\documentclass[12pt]{article}
\usepackage{a4wide,epsfig,amsmath,amssymb,cite,scalefnt}
\usepackage{feynarts}
\usepackage{color}

\newcommand{\abbrev}{\scalefont{.9}}
\newcommand{\drbar}{$\overline{\mbox{\abbrev DR}}$}
\newcommand{\msbar}{$\overline{\mbox{\abbrev MS}}$}

\definecolor{NZGreen}{RGB}{0,225,0}

\begin{document}

%- }}}
%- {{{ title and abstract

\title{
${\cal O}(\alpha_s \alpha_t)$  (non)decoupling effects within the top-sector
  of the MSSM}
\author{ \small 
   L. Mihaila and N. Zerf} 
\date{%
{\small\it   Institut f{\"u}r Theoretische Physik,  University of
  Heidelberg, 69117 Heidelberg, Germany}\\
}

%\date{}

\maketitle

\thispagestyle{empty}

\begin{abstract}
In this paper we compute the $\mathcal{O}(\alpha_s \alpha_t)$ threshold
corrections to the running strong coupling constant, the top-Yukawa coupling
and the top-quark mass within the MSSM.  These parameters present a
non-decoupling behaviour with the supersymmetry scale $M_{\rm SUSY}$.  Our numerical
analysis shows that the mixed QCD-Yukawa corrections can amount to few GeV for
the running top-quark mass and range at the percent level for the top-Yukawa
coupling. 
\medskip

\noindent
PACS numbers: 11.30.Pb, 12.38.-t, 12.38.Bx, 12.10.Kt

\end{abstract}

%%%%%%%%%%%%%%%%%%%%%%%%%%%%%%%%%%

\section{Motivation}
After the end of the LHC run I and the start of the LHC run II, the direct
searches for supersymmetric particles remained unsuccessful, thus increasing the
exclusion bounds for the supersymmetric mass scale towards the TeV
range~\cite{Atlas:2016}. For the time being, it seems that if supersymmetry
is realised in high energy physics the 
most likely scenario is the one where all SUSY particles are much heavier than the
SM ones. For such a  mass hierarchy (also called  the 
``decoupling limit''~\cite{Gunion:1989we}), it was shown~\cite{Dobado:2000pw}
that the effects of the SUSY particles on 
many physical observables in the gauge and Higgs sectors  scales like $M_{\rm EW}^2/M_{\rm SUSY}^2$.
Therefore, for very heavy SUSY scales, deviations from the SM predictions for these
observables will be very challenging to  detect.
 In such a case,  additional efforts both in theory and experiment are
required to identify physical observables for which the power suppressed behaviour does not
occur, or the decoupling limit is delayed by parametric enhancements like in the
presence of large $\tan\beta$ values~\cite{Hall:1993gn}.  Prominent
examples for this  class of  observables are the mass of the lightest Higgs boson and  its
self coupling in SUSY theories that receive  radiative corrections increasing
with the  SUSY mass scale and mixing parameters. 

In this paper we study the
behaviour of the running top  quark mass and  Yukawa coupling   in the
decoupling limit of the MSSM. Although, the running parameters are not themselves
observables, they can be related in perturbation theory with more physical
parameters like for example effective charges in the MOM
scheme~\cite{Celmaster:1979km} or  
short-distance masses in MSR~\cite{Hoang:2008yj}.\footnote{Nevertheless, the explicit
conversion of the running parameters to physical observables  is beyond the
scope of this paper.}
Under the decoupling limit we understand that
all the  superpartners, including the additional Higgs particles $A,
H^0,H^{\pm}$, are much heavier than the SM particles. We assume here that  the
relation  $M_{\rm SUSY},M_{\rm A} \gg
M_{\rm EW}$ holds. The case for an intermediate Higgs sector ( $M_{\rm SUSY}\gg M_{\rm A} \gg M_{\rm
  EW}$), {\it i.e } at intermediate energies the 2HDM is at work,
  was studied  in detail in the
  literature~\cite{Hall:1993gn}. Here, we consider the 
  case  for which the lightest Higgs boson behaves  SM
  like and no additional particles have masses at intermediate energy scales.\\ 
 As
  will be shown in the next sections, the  non-decoupling effects for the running quark
  masses and Yukawa couplings increase logarithmically with the SUSY mass
  scale and have a polynomial dependence with maximal degree three on the mixing parameters in the squark
  sectors. From a phenomenological point of view, the non-decoupling 
  effects are usually few times larger than the current accuracy on the bottom
  and top quark
  mass
  determinations~\cite{Chetyrkin:2009fv,Moch:2014tta,Tevatron:2014cka}. For
  Yukawa couplings the effects are  in the range of 
  precision expected for a  100~TeV collider~\cite{Plehn:2015cta}. 
 In order to keep the theoretical uncertainties well below the experimental
 accuracy, we consider here the two-loop radiative corrections of order ${\cal
   O}(\alpha_s^2,\alpha_s\alpha_t)$ induced by the strong
  and the top-Yukawa couplings, and neglect the bottom-Yukawa
 coupling except for scenarios with large $\tan\beta$ values.  Here, we
 concentrate on the theoretical aspects and give some details of the 
 calculation. 

 The outline of this paper  is  as follows: in the next section we describe briefly
 the computational framework and introduce our notation. In section 3 we provide analytical results
for the decoupling coefficients and discuss their phenomenological
implications in section 4. Our findings are summarized in section 5.

\section{Framework}

In the following, we assume that all the SM particles are
much lighter than any super partner. In this case, the physical phenomena at
low energies can be described with an 
effective theory, called the SMEFT (see Ref.~\cite{deFlorian:2016spz} for a recent 
review), containing six quark flavours and the light Higgs. 
At leading order in  the heavy masses, the effective Lagrangian ${\cal L}_{\rm
  eff}$ can be written as a linear combination of 
  operators  constructed from the light degrees of freedom. We adopt here a
``top-down'' approach, in the sense that the Wilson coefficients of the operators
  will be computed in the full theory, {\it i.e} MSSM for our specific calculation.
For the study of the
  running quark masses and Yukawa couplings up to  ${\cal
    O}(\alpha_s^2,\alpha_s\alpha_t)$ it is sufficient to study the following
  three operators ~\cite{Spiridonov:1984}
\begin{eqnarray}
{\cal L_{\rm eff} }= {\cal L}_{\rm SM}^{(6)}+ {\cal L}^h_{\rm eff} + \ldots\,;\quad
{\cal L}^h_{\rm eff}  = -\frac{h^{(0)}}{v^{(0)}}\left[
C_1^0{\cal O}_1^0 + \sum_{q=u,d,\ldots}\left( C_{2q}^0{\cal O}_{2q}^0
 + C_{3q}^0{\cal O}_{3q}^0\right)+\ldots
\right]\,.
\label{eq::eft}
\end{eqnarray}
 Here $ {\cal L}_{\rm SM}^{(6)}$ denotes the usual SM Lagrangian  with six active quark
 flavours  but without the Yukawa sector. The latter   is considered explicitely in
 ${\cal L}^h_{\rm eff}$, that collects all interactions with just one Higgs
 field. The ellipsis stand for the remaining higher-dimensional operators that we do
 not study here.
  The Wilson coefficients  $C_i\,, i=1,2q,3q$, parameterize the
 effects of the heavy particles and 
 the superscript
 $0$ labels bare quantities. The three
 operators  relevant in our computation are defined as
\begin{eqnarray}
{\cal O}_1^0 &=& (G_{\mu,\nu}^{0,\prime,a})^2\,,\nonumber\\
{\cal O}_{2q}^0 &=& m_q^{0,\prime}\bar{q}^{0,\prime}
q^{0,\prime}\,,\nonumber\\
{\cal O}_{3q}^0 &=& \bar{q}^{0,\prime}(i\,/\!\!\!\! D^{0,\prime}
-m_q^{0,\prime})q^{0,\prime}\,,
\label{eq::ops}
\end{eqnarray}
where $G_{\mu,\nu}^{0,\prime,a}$ and $ D_{\mu}^{0,\prime}$ are the
gluon field strength tensor and the covariant derivative for quark fields, and 
the primes label the quantities in the effective theory. 
 The operator ${\cal O}_{3q}$
 vanishes by the fermionic equation
 of motion and it will not contribute to physical observables.
 So, the last term in Eq.~(\ref{eq::eft})  might be omitted,
 once the coefficients $C_1^0, C_{2q}^0$ are determined. For an easier
 comparison with results already available in the literature, we stick to the
 normalization of the operators  as stated above. The coefficients $C_1$ and
 $C_{2q}$ of the physical operators  can be related with the definitions used
 in Ref.~\cite{deFlorian:2016spz}, via the redefinitions $C_1=c_g\,( \pi
   \alpha_s v^2)/\Lambda^2$ and $C_{2q}=1+c_q\, (16 \pi \alpha_s
   v^2)/\Lambda^2$, where $\Lambda$ denotes the scale of new physics.  For
   intermediate scales $\Lambda$  the effects of new physics are embedded in
   the coefficients of the operators of dimension $n$, with $n\ge 4$. 
In the
   decoupling limit ($\Lambda\to\infty$)  (that we assume here) $C_1\to 0$ and
   $C_{2q}\to 1$ and the higher-dimensional operators decouple. In the next
   section we will verify through an explicit calculation that these relations
   hold at one and two loops in the top sector. 

\begin{figure}[t]
  \begin{center}
    \begin{tabular}{ccc}
\unitlength=1.bp%
\epsfig{file=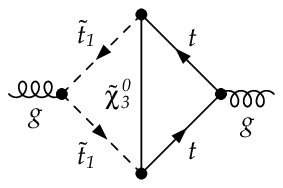,bbllx=80pt,bblly=600pt,bburx=170pt,bbury=670pt,width=0.2\textwidth,clip}&\hspace*{-0.8cm}
\epsfig{file=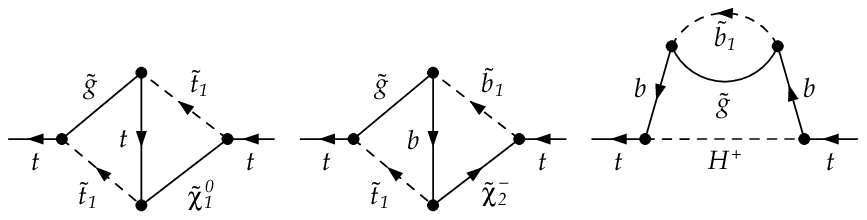,bbllx=80pt,bblly=600pt,bburx=340pt,bbury=670pt,width=0.6\textwidth,clip}&\hspace*{-0.8cm}
\epsfig{file=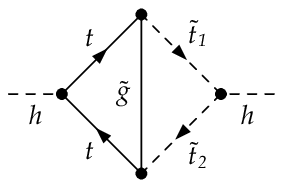,bbllx=80pt,bblly=600pt,bburx=170pt,bbury=670pt,width=0.2\textwidth,clip}
    \end{tabular}
    \parbox{14.cm}{
      \caption[]{\label{fig::prop}\sloppy Sample diagrams
        contributing to $\zeta_3$, $\zeta_{2L,2R}$, $\zeta_{m_t}$ and $\zeta_h$ with
        gluons ($g$),  top quarks ($t$), and light Higgs ($h$) as external particles.}}
  \end{center}
\end{figure}

Furthermore, one has to perform a redefinition of the parameters and fields
present in the EFT. In mass independent renormalization schemes the Appelquist-Carrazone decoupling
theorem~\cite{Appelquist:1974tg} does not hold and the Green functions usually
depend on the heavy masses and mixing parameters present in the full theory. 
In order to avoid potentially large logarithms and power corrections  one has to
decouple the heavy particles from the theory~\cite{Bernreuther:1981sg}.  As
a consequence, one obtains different Lagrangian parameters in the various
energy regimes.  They are connected by the so-called decoupling or
matching relations.
The relations between the quantities defined in the full (MSSM)
and effective theories (SMEFT) that are of interest for our study, 
are given by
\begin{eqnarray}
G_{\mu,\nu}^{0,\prime,a}&=&(\zeta_3^{(0)})^{1/2}
G_{\mu,\nu}^{0,a}\,,\nonumber\\
c^{0,\prime}&=& (\zeta_{3c}^{(0)})^{1/2} c^{0}\,,\nonumber\\
q_L^{0,\prime}&=& (\zeta_{2L}^{(0)})^{1/2} q_L^{0}\,,\nonumber\\
q_R^{0,\prime}&=& (\zeta_{2R}^{(0)})^{1/2} q_L^{0}\,,\nonumber\\
g_s^{0,\prime}&=& \zeta_g^{(0)} g_s^{0}\,,\nonumber\\
m_q^{0,\prime}&=&\zeta_{m_q}^{(0)}m_q^{0}\,,\nonumber\\
h^{0,\prime}&=& (\zeta_h^{(0)})^{1/2} h^{0}\,,\nonumber\\
y_q^{0,\prime}&=& \zeta_{y_q}^{(0)} y_q^{0} f_q(\beta)\,,
\label{eq::dec}
\end{eqnarray}
where $g_s=\sqrt{4\pi\alpha_s}$ is the strong coupling,
$y_q=\sqrt{4\pi\alpha_q}$ is the Yukawa coupling, with $q=b,t$ for the present
calculation, and $f_t(\beta)=\sin\beta$ and $f_b(\beta)=\cos\beta$ with
$\tan\beta$ defined as the ratio of the vacuum expectation values of the two
Higgs doublets. $h$ stands for the field of the lightest neutral Higgs boson. The coefficients
$\zeta_3^{(0)}\,,\zeta_{2L}^{(0)}\,,\zeta_{2R}^{(0)}\,,\zeta_g^{(0)}\,,\zeta_m^{(0)}\,,\zeta_h^{(0)}\,,\zeta_{y_q}^{(0)}$ are the bare
decoupling coefficients. They may be computed from the transverse part
 of the gluon polarization  function,  the vector, axial-vector and scalar
part of the quark self-energy, and the Higgs self-energy   via~\cite{Chetyrkin:1997un}
\begin{eqnarray}
\zeta_3^{(0)}&=&1+\Pi^{0,h}_g(0)\,,\nonumber\\
\zeta_{2L}^{(0)}&=&1+\Sigma_v^{0,h}(0)-\Sigma_A^{0,h}(0)\,,\nonumber\\
\zeta_{2R}^{(0)}&=&1+\Sigma_v^{0,h}(0)+\Sigma_A^{0,h}(0)\,,\nonumber\\
\zeta_h^{(0)}&=&1+\Pi_h^{0,h}(0)\,,\nonumber\\
\zeta_{m_q}^{(0)}&=&\frac{1-\Sigma_s^{0,h}(0)}{\sqrt{(1+\Sigma_v^{0,h}(0))^2-\Sigma_A^{0,h}(0)^2}}\,. 
\label{eq::letse}
\end{eqnarray}
The pseudo-scalar part of the quark self-energy vanishes both in the SM and
the MSSM 
through two loops, that renders the equation for $\zeta_m^{(0)}$ in the
reduced form given above. Decoupling coefficients are renormalized in a
similar manner with the  associated parameters and
fields~\cite{Chetyrkin:1997un}. The superscript $h$ indicates that at least
one heavy particle is present in the diagrams. Sample Feynman diagrams are
shown in Fig.~\ref{fig::prop}.

\begin{figure}[t]
  \begin{center}
    \begin{tabular}{c}
\unitlength=1.bp%
\epsfig{file=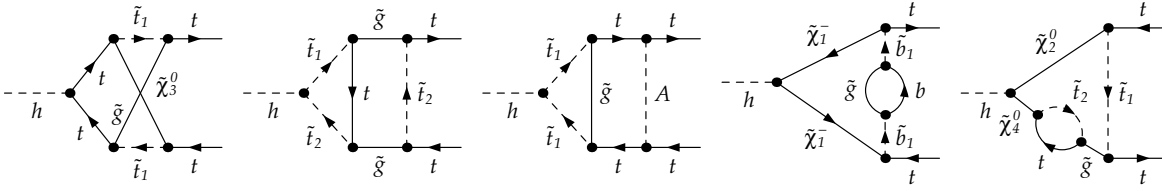,bbllx=80pt,bblly=600pt,bburx=500pt,bbury=670pt,width=0.99\textwidth,clip}
    \end{tabular}
    \parbox{14.cm}{
      \caption[]{\label{fig::vertex}\sloppy Sample diagrams
        contributing to $\zeta_{hqq}$.}}
  \end{center}
\end{figure}

For the derivation of the coefficient $\zeta_g^{(0)}$ and $\zeta_{y_q}^{(0)}$
one has to 
consider in addition vertices involving the strong and Yukawa couplings. For
example, we chose,
\begin{eqnarray}
\zeta_{g}^{(0)}&=&\frac{\zeta_{gcc}^{(0)}}{\sqrt{\zeta_3^{(0)}\zeta_{3c}^{(0)}}}\,,\quad\mbox{with}\quad
\zeta_{gcc}^{(0)}=1+\Gamma_{gcc}^{0,h}(0,0)\nonumber\\
\zeta_{y_q}^{(0)}&=&\frac{\zeta_{hqq}^{(0)}}{\sqrt{\zeta_h^{(0)}\zeta_{2L}^{(0)}\zeta_{2R}^{(0)}}}\,,\quad\mbox{with}\quad
\zeta_{hqq}^{(0)}=1+\Gamma_{hqq}^{0,h}(0,0)\,.
\label{eq::letg}
\end{eqnarray}
Here $\Gamma_{gcc}^{0,h}(0,0)$ and $\Gamma_{htt}^{0,h}(0,0)$ denote the
one-particle-irreducible vertex functions with vanishing momenta flowing
through the external legs. Sample Feynman diagrams  for the two vertices can
be seen in Fig.~\ref{fig::vertex}.
The decoupling coefficients are independent of the momentum transfer, so
that they can be evaluated at vanishing external momenta.
%The superscript $h$ indicates that only diagrams
%containing at least one heavy particle inside the loops contribute. 

 The  two-loop 
 SUSY-QCD contributions to $\zeta_g$ and $\zeta_{m_b}$ for
 non-vanishing top-quark mass were computed
 in ~\cite{Harlander:2005wm}. The SUSY-QCD contributions
for  $\zeta_g$, $\zeta_{m_t}$ and $\zeta_{y_t}$ in the same  limit
as in the present paper are available from Ref~\cite{Kunz:2014gya}. In addition,
two-loop Yukawa contributions to  $\zeta_{m_b}$ and $\zeta_{m_\tau}$ were
computed in the limit of  heavy  Higgses (including the lightest one) in
Ref.~\cite{Bednyakov:2009wt}. Two-loop SUSY-QCD and dominant Yukawa
contributions for the scalar part of the bottom-quark self energy for the case 
of an intermediate Higgs mass range ({\it i.e.} the EFT is the 2HDM) are 
available from Ref.~\cite{Noth:2010jy}.  The two-loop SUSY-QCD corrections to
the 
vertex functions in the above mentioned limit  can be found in
Ref.~\cite{Mihaila:2010mp}. ${\cal O}(\alpha_s\alpha_t)$ corrections to the
decouplin g
coefficients  $\zeta_g$, $\zeta_{m_t}$ and $\zeta_{y_t}$ 
for a specific mass hierarchy have been computed in~\cite{kunz:2015}.
% The present calculation is devoted to the mixed
%Yukawa-SQCD contributions to the decoupling coefficients for the parameters
%introduced in Eqs.~\ref{eq::letse,eq::letg}, in the decoupling limit of the MSSM.
 
In the present paper we consider the following tree level  mass hierarchies
\begin{eqnarray}
M_{\rm
  SUSY}\,;M_{\tilde{Q}}\,;M_{\tilde{D}}\,;M_{\tilde{U}}\,;M_{\tilde{g}}\,;\mu\,
&\gg& M_{t}\,;M_{Z}\,\nonumber\\
M_{\rm SUSY}\,;M_{H}\simeq M_{H^\pm}\simeq M_{A} &\gg& M_Z\,;M_h\,,
\label{eq:dec}
\end{eqnarray}
where the semicolon between the mass parameters indicates that they are of
similar size but not necessarily equal  and $\simeq$ means that the masses are
equal up to corrections of ${\cal O}(M_Z^2/M_A^2)$, that we neglect in the
calculation. We introduce $M_{\rm SUSY}$ as a generic heavy mass scale of the
order of the supersymmetric particle masses, that will be used in the
following discussions. Precisely, for a degenerate supersymmetric mass spectrum
it becomes equal to this mass. For a non degenerate spectrum it denotes the
mass scale at which the supersymmetric particles become active.  The rest of the parameters in the above
inequalities denotes the usual soft SUSY breaking parameters in the squark
sector.   Beyond tree level, it has been shown~\cite{Haber:1994mt} that this mass
 hierarchy is stable under radiative corrections.  A similar observation holds also for 
the strongly interacting sector. Let us mention   that by expanding in
inverse powers of $M_A$, we get the following relations for the mixing angles in
the Higgs sector
\begin{eqnarray}
&&\frac{\cos\alpha}{\sin\beta} = 1+{\cal
  O}\left(\frac{M_Z^2}{M_A^2}\right)\,,\quad
%\nonumber\\
-\frac{\sin\alpha}{\cos\beta} = 1+{\cal
  O}\left(\frac{M_Z^2}{M_A^2}\right)\,\nonumber\\
&&\sin(\beta-\alpha) = 1+{\cal
  O}\left(\frac{M_Z^4}{M_A^4}\right)\,.
\label{eq:allig}
\end{eqnarray}
These relations  parameterize the modifications of the lightest
Higgs boson couplings to up-type  ($\cos\alpha/\sin\beta$) and down-type
($\sin\alpha/\cos\beta$) quarks and to gauge bosons ($\sin(\beta-\alpha)$) in
the MSSM as compared to the SM.
Thus, in the decoupling limit the tree-level couplings of the lightest Higgs boson tend to their SM values, as assumed in
Eq.~(\ref{eq::eft}). These relations  hold also  at one and two loops and
translates into the equality $C_{2t}=1$ (in the decoupling limit). This will
be  discussed in more detail  below Eq.~(\ref{eq::coeff}) in the next section. 

In the calculation, we take into account  the complete one-loop contributions, 
including genuine electroweak diagrams. At the two-loop order, we neglect
 electroweak effects (also known as the gaugeless limit). The last
approximation is motivated by the small
size of the one-loop electroweak effects on the considered parameters.
Nevertheless, for the one-loop calculation we assume the following mass
hierarchy in the EW sector
\begin{eqnarray}
M_{\rm  SUSY}\,;M_1\,;M_2\,;\mu\gg M_Z\,;M_W\,.
\end{eqnarray}
In this limit, the mass matrices for charginos and neutralinos become
diagonal and there is no mixing between higgsinos and electroweakinos. Thus, at the
two-loop order  we will practically take into account  only  diagrams with higgsinos.

 In our setup, the Feynman diagrams are generated with the program {\tt QGRAF}~\cite{Nogueira:1991ex}, and further
processed with {\tt q2e} and {\tt
  exp}~\cite{Seidensticker:1999bb,Davydychev:1992mt}. In the approximation we
worked out,  the propagator and 
vertex topologies can be reduced to  two-loop tadpoles~\cite{Davydychev:1992mt}.
The reduction of various vacuum integrals to the
master integral was performed by a self-written {\tt FORM}~\cite{Vermaseren:2000nd}  routine.

For the regularization of the SMEFT we used dimensional regularization (DREG) and for
the underlying supersymmetric theory the dimensional reduction (DRED). Technically,
we implemented DRED with the help of $\varepsilon$-scalars for the $\gamma, W,
Z, g$ and assigned them heavy masses~\cite{Harlander:2005wm}. There are
various ways to perform the explicit calculation. One can   in a first step  decouple  the physical 
supersymmetric particles using DRED both for the underlying theory as well as
for the  SMEFT. In a second step, one has to change the regularization scheme
from DRED to DREG.\footnote{For a detailed discussion see for example
  Ref.~\cite{Harlander:2007wh}.} However, in the second step a tower of 
new  $\varepsilon$-scalar couplings, the so called evanescent couplings, have 
 to be taken into account. One can avoid this technical complication using
 the fact that the change of regularization scheme from DRED to DREG is
 equivalent to integrating out the   $\varepsilon$-scalars from the theory.
 This last operation  means that one can also  apply the decoupling procedure to the
 $\varepsilon$-scalars. Because the physical predictions do not depend on
 $\varepsilon$-scalar masses, we are free to choose them to be heavy and 
 decouple them  together with the heavy supersymmetric particles.
 In this way, we can perform the change of
regularization scheme together with  the decoupling procedure. As a consistency
check, we verified through two loops that the decoupling coefficients do not depend on the
unphysical mass parameters associated with the  $\varepsilon$-scalars, as
expected. In order to get rid of the unphysical 
$\varepsilon$-scalar masses, finite shifts in the one-loop renormalization
constants for squark masses have to be introduced~\cite{Jack:1994rk}.

For the renormalization of the UV divergences, we employed the minimal
subtraction scheme \drbar{}$^\prime$~\cite{Jack:1994rk} for the  coupling constants, quark masses, squark
masses and mixing angles and the mass of the $\varepsilon$-scalar
associated with the gluon. Explicitly, we keep only the divergent terms in the
renormalization constants for all the parameters, except for the squark masses
that get  finite shifts proportional with the mass of the $\varepsilon$-scalar
associated with the gluon. The finite terms are chosen such that the 
$\varepsilon$-scalar mass $M_{\varepsilon}$ decouples from the system of the renormalization
group equations for the squark masses.
Explicit formulae for the
finite shifts, the renormalization constants for the $\varepsilon$-scalar mass
and a detailed description of the method up to three loops in 
SUSY-QCD can be found in Ref.~\cite{Hermann:2011ha}. 
The renormalization constants for the parameters $X_t=A_t-\mu/\tan\beta$ and
$X_b=A_b-\mu\tan\beta$ can be derived by the chain rule from the definition of the mixing
angles. Let us mention that the contributions proportional to the top- and
bottom-quark masses  in the renormalization constants for the squark masses
and mixing angles do not contribute to the decoupling coefficients. However,
such terms have to be kept in the derivation of the  renormalization constants
for $X_t$ and $X_b$. Explicit one-loop formulae for the renormalization
constants for the quark and squark masses and mixing angle can be found in
Ref.~\cite{Bednyakov:2009wt}.  For the limit we are interested in here, one
has to set  the Goldstone boson masses to zero and apply  Eqs.~(\ref{eq:allig}).
Furthermore, we renormalized the tadpole contributions both in the SMEFT and
the underlying theory following the prescription introduced in
Ref.~\cite{Sirlin:1985ux}. It generically requires that
\begin{eqnarray}
\delta t_i^{(l)} +T_i^{(l)}=0\,,
\end{eqnarray}
at all orders in perturbation theory. Here, $\delta t_i$ denotes the $l$-loop
tadpole counterterm and $T_i$ the  $l$-loop tadpole diagrams, and $ i =1, 2$
sums over the Higgs doublets. In this scheme, no explicit tadpole
contributions have to be taken into account in the calculation.

\section{Analytical results}
The one-loop results for the decoupling coefficients of the gauge 
and top-Yukawa couplings in the MSSM can be found in
Ref.~\cite{Bagnaschi:2014rsa}. For completeness we give in the Appendix the
one-loop contributions to $\zeta_{m_t}$.

The analytic formulae for the two-loop results are too long to be presented
here. They will be used for the  numerical analysis in the next section. 
 However, for a better understanding of the numerical effects, we provide
 below the results for two simplified  mass hierarchies. For brevity we display only the
 contributions of
 $\mathcal{O}(\alpha_s\,,\alpha_t\,,\alpha_s^2,\,,\alpha_s\alpha_t)$. The
 complete results are submitted in electronic form together with the paper.

 A) degenerate mass spectrum:\\
 $M_{\rm SUSY}=M_{\tilde{Q}_i}=M_{\tilde{D}_i}=M_{\tilde{U}_i}=M_{\tilde{g}}=M_A=\mu\gg
 m_t\,;M_Z\,;M_h$
\begin{align}
\zeta_{\alpha_s}=& 1+ a_s\bigg[
C_A\left(-\frac{1}{3} - \frac{2}{3} L_{S}\right) - \frac{4}{3} I_{2R} n_g  L_{S}
\bigg]+a_s^2 \bigg[
C_A^2\left(-\frac{5}{18} - \frac{8}{9} L_{S} + \frac{4}{9} L_{S}^2\right) 
\nonumber\\&+
C_R I_{2R} n_g\left(\frac{26}{3} - \frac{8}{3} L_{S}\right) + 
 C_A I_{2R} n_g \left(-\frac{4}{9} + \frac{8}{9} L_S + \frac{16}{9}
 L_S^2\right) + \frac{16}{9} I_{2R}^2 n_g ^2 L_S^2
\bigg]\nonumber\\&+\displaybreak[1]
a_s a_t I_{2R}\bigg\{s_{\beta}^2\bigg[\frac{1}{3} - \frac{8}{3} L_S +
  \left(\frac{10}{9} - 8 S_2\right)
  \tilde{X}_t^2\bigg]+\frac{c_{\beta}}{s_{\beta}}
\left(
-\frac{32}{9} - \frac{8}{3} L_S + 16 S_2
\right)\tilde{X}_t
\nonumber\\&\displaybreak[1]
+\frac{1}{s_{\beta}^2}\left(-\frac{16}{9} - \frac{4}{3} L_S
+ 8 S_2\right)+\bigg[
-\frac{4}{3} + \frac{20}{3} L_S + \left(-\frac{16}{9} - \frac{4}{3} L_S + 8 S_2\right)\tilde{X}_t^2
\bigg]
\bigg\}\,,
\label{eq:zetas1}
\end{align}

\begin{align}
\zeta_{\alpha_t}=& 1+ 
  a_s C_R\left(2  -2 L_{S} -2  \tilde{X}_t\right)
 +  a_t \bigg[-\frac{3}{2} 
 L_{S}
 - c_{\beta}^2 \left( \frac{3}{4} 
  +\frac{3}{2} L_{S}
\right) 
  -\frac{1}{2} s_{\beta}^2 
 \tilde{X}_t^2
\bigg]\nonumber\\&\displaybreak[1]
+ a_s^2 \bigg\{
C_R^2 \bigg[\frac{119}{12} 
  -\frac{5}{3} L_{S}
 + 2 L_{S}^2 
 + \left( -\frac{4}{3} 
 + \frac{28}{3} L_{S}
\right) \tilde{X}_t
 + 
 \tilde{X}_t^2
\bigg]
\nonumber\\&\displaybreak[1] + 
 C_A C_R \bigg[\frac{217}{36} 
  -\frac{25}{3} L_{S}
  - L_{S}^2
 + \left( -\frac{16}{3} 
  -8 L_{S}
\right) \tilde{X}_t
\bigg] \nonumber\\&+ \displaybreak[1]
C_R I_{2R} n_g\bigg[\frac{170}{9} 
  -16 L_{S}
 + 8 L_{S}^2
 + \left( -\frac{16}{3} 
 + \frac{32}{3} L_{S}
\right) \tilde{X}_t
\bigg]
 \bigg\} 
\nonumber\\&+  a_s a_t C_R\bigg\{
s_{\beta}^2 \bigg[-\frac{67}{4} + \frac{61}{2} L_{S} - \frac{9}{2} L_{S}^2  +
  \frac{459}{4} S_2+ 9 \zeta(2) + \left(\frac{11}{2} - 6 L_{S} + \frac{27}{2} S_2
  \right)\tilde{X}_t 
\nonumber\\&\displaybreak[1]
+ \left(-\frac{67}{12} + L_{S} - 3 S_2\right)\tilde{X}_t^2 + 
  \left(\frac{8}{3} - 12 S_2\right)\tilde{X}_t^3 
\bigg]
+\frac{c_{\beta}}{s_{\beta}}
\bigg[
6 + 6 L_{S}- \frac{27}{2} S_2
\nonumber\\&
 + \left(-\frac{13}{3} - 4 L_{S}+ 15
S_2\right)\tilde{X}_t + \left(-\frac{16}{3} - 4 L_{S} + 24 S_2\right)\tilde{X}_t^2
\bigg]
+\frac{1}{s_{\beta}^2}
\bigg[-\frac{13}{6} - 2 L_{S} 
\nonumber\\&\displaybreak[1]
+ 12 S_2 
+ \left(-\frac{8}{3} - 2 L_{S} + 12 S_2\right)\tilde{X}_t
\bigg]+\bigg[
-\frac{9}{4} - \frac{11}{2} L_{S} - \frac{459}{4} S_2 - 9 \zeta(2)
 \nonumber\\&
+ \left(\frac{23}{2} + 16 L_{S}  - \frac{27}{2} S_2\right)\tilde{X}_t
 +
\left(-\frac{13}{6} - 2 L_{S} + 3 S_2\right)\tilde{X}_t^2 
\nonumber\\&
+ 
 \left(-\frac{8}{3} - 2 L_{S} + 12 S_2\right)\tilde{X}_t^3 
\bigg]
\bigg\}\,,\displaybreak[1]
\label{eq:zetat1}
\end{align}

\begin{align}
\zeta_{m_t}=&1+
a_s C_R \left(1 
  - L_{S} -
 \tilde{X}_t
\right)+
 a_t \bigg[s_{\beta}^2 \left(\frac{3}{8} 
 + \frac{3}{4} L_{S}
\right) 
 -
 \left( \frac{3}{8} 
  +\frac{3}{2} L_{S}
\right) 
\bigg]\nonumber\\&\displaybreak[1]
+ 
a_s^2\bigg\{
 C_R^2 \bigg[\frac{107}{24} 
 + \frac{1}{6} L_{S}
 + \frac{1}{2} L_{S}^2
 + \left(\frac{1}{3} 
 + \frac{11}{3} L_{S}
\right) \tilde{X}_t
\bigg]
\nonumber\\&
 + 
 C_A C_R \bigg[\left(\frac{217}{72} 
  -\frac{25}{6} L_{S}
  -\frac{1}{2} L_{S}^2
\right) 
 + \left( -\frac{8}{3} 
  -4 L_{S}
\right) \tilde{X}_t
\bigg]
 \nonumber\\& \displaybreak[1]
+
C_R I_{2R}n_g \bigg[\left(\frac{85}{9} 
  -8 L_{S}
 + 4 L_{S}^2
\right) 
 + \left( -\frac{8}{3} 
 + \frac{16}{3} L_{S}
\right) \tilde{X}_t
\bigg]
\bigg\}
\nonumber\\&
+  a_s a_t C_R \bigg\{
 s_{\beta}^2 \bigg[-\frac{37}{8}+ \frac{83}{8} L_{S}+\frac{459 S_2}{8}+\frac{9 \zeta(2)}{2}
 + \left(-\frac{7}{8}-\frac{5}{4} L_{S}+\frac{27}{4} S_2
\right) \tilde{X}_t
\nonumber\\&
 - \left(\frac{13}{24}+\frac{3}{2} S_2\right) 
 \tilde{X}_t^2
 + \left(\frac{5}{6}-6 S_2\right) 
 \tilde{X}_t^3
\bigg]
 + \frac{c_{\beta}}{s_{\beta}} \bigg[3+ 3 L_{S}-\frac{27}{4} S_2
\nonumber\\&\displaybreak[1]
 +
\left(-\frac{13}{6}-2 L_{S}+\frac{15}{2} S_2 
  \right) \tilde{X}_t
 + \left(-\frac{8}{3} 
  -2 L_{S}+12 S_2
\right) \tilde{X}_t^2
\bigg] \nonumber\\& 
+ \frac{1}{s_{\beta}^{2}} \bigg[
\left(-\frac{13}{12}- L_{S}+6 S_2  \right) 
 - \left(\frac{4}{3}+ L_{S}-6 S_2 
\right) \tilde{X}_t
\bigg]\nonumber\\&
+ 
 \bigg[-\frac{3}{4}-\frac{13}{8} L_{S}
  -\frac{3}{2} L_{S}^2-\frac{459 S_2}{8}-\frac{9 \zeta(2)}{2}
+ \left(\frac{43}{8}+ \frac{13}{2} L_{S}-\frac{27}{4} S_2 
\right) \tilde{X}_t
\nonumber\\&
 + \left(-\frac{13}{12} - L_{S}+\frac{3}{2} S_2
\right) \tilde{X}_t^2
 + \left(-\frac{4}{3}+6 S_2 - L_{S}
\right) \tilde{X}_t^3
\bigg]
\bigg\}\,,  
\displaybreak[1]
\label{eq:zetam1}
\end{align}

where $a_i=\alpha_i/(4\pi)$, $S_2 =
\frac{4}{9 \sqrt{3}}{\rm Cl}_2(\frac{\pi}{3}) \simeq 0.260434$ and
$\zeta(2)=\frac{\pi^2}{6}$. 
 $C_A\,,C_R$ are the
$SU(3)$ Casimir invariants for the adjoint and fundamental representations,
 $I_{2R}=1/2$, and $n_g=3$ denotes the number of generations. We have used
further the abbreviations $\tilde{X}_t=\frac{X_t}{M_{\rm SUSY}}$,
$L_S=\ln\left(\frac{\mu_{\rm dec}^2}{M_{\rm SUSY}^2}\right)$, 
$c_{\beta}=\cos\beta$ and $s_{\beta}=\sin\beta$, where $\mu_{\rm dec}$ denotes
the decoupling scale at which the heavy degrees of freedom are integrated
out. Its  precise value is not fixed by theory. Since the explicit dependence
on this matching scale is logarithmic, it is natural to choose $\mu_{\rm
  dec}\approx M_{\rm SUSY}$. The dependence of theoretical predictions on the
variation of the matching
scale around this intuitive value can be used as an estimate of the
theoretical uncertainty. Reduction of this uncertainty can only be
achieved by means of higher order calculations.
\\

B) light higgsino masses:\\
 $M_{\rm SUSY}=M_{\tilde{Q}_i}=M_{\tilde{D}_i}=M_{\tilde{U}_i}=M_{\tilde{g}}=M_A\gg\mu\gg
 m_t\,;M_Z\,;M_h$
\begin{align}
\zeta_{\alpha_s}=& 1+ a_s\bigg[
C_A\left(-\frac{1}{3} - \frac{2}{3} L_{S}\right) - \frac{4}{3} I_{2R} n_g  L_{S}
\bigg]+a_s^2 \bigg[
C_A^2\left(-\frac{5}{18} - \frac{8}{9} L_{S} + \frac{4}{9} L_{S}^2\right) 
\nonumber\\&+
C_R I_{2R} n_g\left(\frac{26}{3} - \frac{8}{3} L_{S}\right) + 
 C_A I_{2R} n_g \left(-\frac{4}{9} + \frac{8}{9} L_S + \frac{16}{9}
 L_S^2\right) + \frac{16}{9} I_{2R}^2 n_g ^2 L_S^2
\bigg]\nonumber\\&+\displaybreak[1]
a_s a_t I_{2R}\bigg\{s_{\beta}^2\bigg[\frac{1}{3} - \frac{8}{3} L_S +
  \left(\frac{10}{9} - 8 S_2\right)
  \tilde{X}_t^2\bigg]+\frac{c_{\beta}}{s_{\beta}}
\left(
-\frac{32}{9} - \frac{8}{3} L_S + 16 S_2
\right)\tilde{X}_t\frac{\mu}{M_{\rm S}}
\nonumber\\&\displaybreak[1]
+\frac{1}{s_{\beta}^2}\left(-\frac{16}{9} - \frac{4}{3} L_S
+ 8 S_2\right)\frac{\mu^2}{M_{\rm S}^2}+\bigg[
-\frac{4}{3} + 4 L_S 
+ \frac{8}{3}\left(1+ L_S\right)\frac{\mu^2}{M_{\rm
    S}^2} 
\nonumber\\&\displaybreak[1]
+ \frac{8}{3}L_{\mu M_S}\frac{\mu^4}{M_{\rm S}^4}
+ \left(-\frac{16}{9} - \frac{4}{3} L_S + 8 S_2\right)\tilde{X}_t^2
\bigg]
\bigg\}+\mathcal{ O}\left(\frac{\mu^6}{M_{\rm S}^6}\right)\,,
\label{eq:zetas2}
\end{align}

\begin{align}
\zeta_{\alpha_t}=& 1+ 
  a_s C_R\left(2  -2 L_{S} -2  \tilde{X}_t\right)
 +  a_t \bigg[-\frac{3}{4}-\frac{3}{2} L_{S}
+\frac{3}{2}\frac{\mu^2}{M_{\rm S}^2}+\frac{3}{2}(1+L_{\mu
  M_S})\frac{\mu^4}{M_{\rm S}^4}
\nonumber\\&
 - c_{\beta}^2 \left( \frac{3}{4} 
  +\frac{3}{2} L_{S}
\right) 
  -\frac{1}{2} s_{\beta}^2 
 \tilde{X}_t^2
\bigg]
+ a_s^2 \bigg\{
C_R^2 \bigg[\frac{119}{12} 
  -\frac{5}{3} L_{S}
 + 2 L_{S}^2 
 + \left( -\frac{4}{3} 
 + \frac{28}{3} L_{S}
\right) \tilde{X}_t
\nonumber\\&
 + 
 \tilde{X}_t^2
\bigg]
 + 
 C_A C_R \bigg[\frac{217}{36} 
  -\frac{25}{3} L_{S}
  - L_{S}^2
 + \left( -\frac{16}{3} 
  -8 L_{S}
\right) \tilde{X}_t
\bigg] \nonumber\\&+ \displaybreak[1]
C_R I_{2R} n_g\bigg[\frac{170}{9} 
  -16 L_{S}
 + 8 L_{S}^2
 + \left( -\frac{16}{3} 
 + \frac{32}{3} L_{S}
\right) \tilde{X}_t
\bigg]
 \bigg\} 
\nonumber\\&+  a_s a_t C_R\bigg\{
s_{\beta}^2 \bigg[-\frac{67}{4} + \frac{61}{2} L_{S} - \frac{9}{2} L_{S}^2  +
  \frac{459}{4} S_2+ 9 \zeta(2) + \left(\frac{11}{2} - 6 L_{S} + \frac{27}{2} S_2
  \right)\tilde{X}_t 
\nonumber\\&\displaybreak[1]
+ \left(-\frac{67}{12} + L_{S} - 3 S_2\right)\tilde{X}_t^2 + 
  \left(\frac{8}{3} - 12 S_2\right)\tilde{X}_t^3 
\bigg]
+\frac{c_{\beta}}{s_{\beta}}
\bigg[
6 + 6 L_{S}- \frac{27}{2} S_2
\nonumber\\&
 + \left(-\frac{13}{3} - 4 L_{S}+ 15
S_2\right)\tilde{X}_t + \left(-\frac{16}{3} - 4 L_{S} + 24 S_2\right)\tilde{X}_t^2
\bigg]\frac{\mu}{M_{\rm S}}
\nonumber\\&\displaybreak[1]
+\frac{1}{s_{\beta}^2}
\bigg[-\frac{13}{6} - 2 L_{S} + 12 S_2 
+ \left(-\frac{8}{3} - 2 L_{S} + 12 S_2\right)\tilde{X}_t
\bigg]\frac{\mu^2}{M_{\rm S}^2}
\nonumber\\&
+\bigg[
-\frac{37}{2} - 9 L_{S} - \frac{459}{4} S_2 +\frac{3}{2} \zeta(2) +
\left(-\frac{49}{2} + L_S + 15\zeta(2)\right)\frac{\mu^2}{M_{\rm S}^2}
\nonumber\\&
+
\left(-\frac{477}{8} - \frac{89}{4} L_S - 9 L_S^2 + \frac{29}{4} L_{\mu} + 9
L_S  L_{\mu} + 24 \zeta(2)\right)\frac{\mu^4}{M_{\rm S}^4}
\nonumber\\& +
\bigg[1 + 12 L_S - \frac{27}{2} S_2\ + 6\zeta(2) + (-5 + 4 L_S +
6\zeta(2))\frac{\mu^2}{M_{\rm S}^2} 
\nonumber\\&
+\left(-\frac{21}{2} + 4 L_{\mu M_S} + 6\zeta(2)\right)\frac{\mu^4}{M_{\rm S}^4} \bigg]\tilde{X}_t
+
\left(-\frac{13}{6} - 2 L_{S} + 3 S_2\right)\tilde{X}_t^2 
\nonumber\\& 
+ 
 \left(-\frac{8}{3} - 2 L_{S} + 12 S_2\right)\tilde{X}_t^3 
\bigg]
\bigg\}+\mathcal{ O}\left(\frac{\mu^6}{M_{\rm S}^6}\right)\,,&\displaybreak[1]
\label{eq:zetat2}
\end{align}
\begin{align}
\zeta_{m_t}=&1+
a_s C_R \left(1 
  - L_{S} -
 \tilde{X}_t
\right)+
 a_t \bigg[s_{\beta}^2 \left(\frac{3}{8} 
 + \frac{3}{4} L_{S}
\right) 
 - \frac{3}{4} 
  -\frac{3}{2} L_{S}+\frac{3}{4}\frac{\mu^2}{M_{\rm S}^2} \nonumber\\&\displaybreak[1]
+ \frac{3}{4}(1 + L_{\mu M_S})\frac{\mu^4}{M_{\rm S}^4}
\bigg] 
+ a_s^2\bigg\{
 C_R^2 \bigg[\frac{107}{24} 
 + \frac{1}{6} L_{S}
 + \frac{1}{2} L_{S}^2
 + \left(\frac{1}{3} 
 + \frac{11}{3} L_{S}
\right) \tilde{X}_t
\bigg] 
\nonumber\\&
+ 
 C_A C_R \bigg[\left(\frac{217}{72} 
  -\frac{25}{6} L_{S}
  -\frac{1}{2} L_{S}^2
\right) 
 + \left( -\frac{8}{3} 
  -4 L_{S}
\right) \tilde{X}_t
\bigg]
 \nonumber\\& \displaybreak[1]
+
C_R I_{2R}n_g \bigg[\left(\frac{85}{9} 
  -8 L_{S}
 + 4 L_{S}^2
\right) 
 + \left( -\frac{8}{3} 
 + \frac{16}{3} L_{S}
\right) \tilde{X}_t
\bigg]
\bigg\}
\nonumber\\&
+  a_s a_t C_R \bigg\{
 s_{\beta}^2 \bigg[-\frac{37}{8}+ \frac{83}{8} L_{S}+\frac{459 S_2}{8}+\frac{9 \zeta(2)}{2}
 + \left(-\frac{7}{8}-\frac{5}{4} L_{S}+\frac{27}{4} S_2
\right) \tilde{X}_t
\nonumber\\&
 - \left(\frac{13}{24}+\frac{3}{2} S_2\right) 
 \tilde{X}_t^2
 + \left(\frac{5}{6}-6 S_2\right) 
 \tilde{X}_t^3
\bigg]
 + \frac{c_{\beta}}{s_{\beta}} \bigg[3+ 3 L_{S}-\frac{27}{4} S_2
\nonumber\\&\displaybreak[1]
 +
\left(-\frac{13}{6}-2 L_{S}+\frac{15}{2} S_2 
  \right) \tilde{X}_t
 + \left(-\frac{8}{3} 
  -2 L_{S}+12 S_2
\right) \tilde{X}_t^2
\bigg]\frac{\mu}{M_{\rm S}} \nonumber\\& 
+ \frac{1}{s_{\beta}^{2}} \bigg[
\left(-\frac{13}{12}- L_{S}+6 S_2  \right) 
 - \left(\frac{4}{3}+ L_{S}-6 S_2 
\right) \tilde{X}_t
\bigg]\frac{\mu^2}{M_{\rm S}^2}\nonumber\\&
+ 
 \bigg[
-\frac{17}{2} - \frac{15}{4} L_S - \frac{3}{2} L_S^2  -\frac{459 S_2}{8} +
\frac{3}{4} \zeta(2) + \left(-13 + \frac{5}{4} L_S +
\frac{15}{2}\zeta(2)\right)\frac{\mu^2}{M_{\rm S}^2} 
\nonumber\\&
+ 
 \left(-\frac{489}{16} - \frac{89}{8} L_S - \frac{15}{4} L_S^2 +
 \frac{35}{8} L_{\mu} + \frac{15}{4} L_S L_{\mu} + 12
 \zeta(2)\right)\frac{\mu^4}{M_{\rm S}^4}
\nonumber\\&\displaybreak[1]
\bigg[-\frac{1}{4} + \frac{9}{2} L_S - \frac{27}{4} S_2 + 3\zeta(2) +
  \left(-\frac{7}{4} + 2 L_S + 3\zeta(2)\right)\frac{\mu^2}{M_{\rm S}^2} 
\nonumber\\&
+ \left(-\frac{9}{2} + \frac{11}{4} L_{\mu M_S} + 3\zeta(2)\right)\frac{\mu^4}{M_{\rm S}^4}
\bigg]\tilde{X}_t
+ \left(-\frac{13}{12} - L_{S}+\frac{3}{2} S_2
\right) \tilde{X}_t^2
\nonumber\\&
 + \left(-\frac{4}{3}+6 S_2 - L_{S}
\right) \tilde{X}_t^3
\bigg]
\bigg\}  +\mathcal{ O}\left(\frac{\mu^6}{M_{\rm S}^6}\right)
\displaybreak[1]\,,
\label{eq:zetam2}
\end{align}
where $L_{\mu M_S}=\ln\left(\frac{\mu^2}{M_{\rm SUSY}^2}\right)$ and $a_s$ and $a_t$ denote the MSSM couplings. 

For the derivation of the above formulae we implemented two methods. In one approach, we
expanded the two-loop analytical results for a general mass spectrum in the
above given mass hierarchies. In the second approach, we asymptotically
expanded the Feynman integrals with the code {\tt exp} and afterwards used
the one-mass scale tadpole integrals from the code {\tt MATAD}~\cite{Steinhauser:2000ry}.  We
found full agreement between the two approaches.

Analytical expressions for the dominant two-loop contributions to 
$\Sigma_{s,b}$ in the case of a degenerate SUSY spectrum, but for intermediate Higgs masses can be found
in~\cite{Noth:2010jy}.   We have computed $\zeta_{s,b}$  also for this mass
hierarchy, and after translating our results into the renormalization and regularization
schemes used in that reference, we found  full analytical agreement.
  
Another interesting consistency check of our calculation is to show that the
Wilson coefficients introduced in Eq.~(\ref{eq::eft}) reach their SM values when the
decoupling limit is applied. Explicitly, it holds
\begin{eqnarray}
C_i&=& C_i^{\rm{SM}}+{\cal O}(\frac{m_t^2}{M_{\rm SUSY}^2},\frac{M_Z^2}{M_{\rm
    SUSY}^2},\frac{M_h^2}{M_{\rm SUSY}^2},)\quad \mbox{with}\quad
i=g\,,2t\,,3t\,.
\label{eq::coeff}
\end{eqnarray} 
The mass suppressed terms will give contributions to Wilson coefficients
associated with dimension 6 or higher operators~\cite{deFlorian:2016spz}, that are beyond the scope of
the present paper. At the one-loop level, it is an easy exercise to apply the
decoupling limit as stated in Eqs.~(\ref{eq:dec}) and (\ref{eq:allig}) to the known results from
Refs.~\cite{Noth:2010jy,Mihaila:2010mp} and get $C_{2t}=1$, $C_g= 0 $ an             
$C_{3t}=0$. The first relation is guaranteed by the alignment
limit Eq.~(\ref{eq:allig}), whereas the other two are due to the limit $m_t\to
0$. At ${\cal O}(\alpha_s^2, \alpha_s\alpha_t, \alpha_s \alpha_b)$ we have
explicitly verified that in the decoupling limit $C_{2t}=1$, that amounts to
show that the corrections to vertex function $\Gamma_{htt}^{h}(0,0)$ and  to
the scalar part of the top-quark selfenergy $\Sigma_s^{h}(0)$ are equal up to a sign. Furthermore,
 the vertex function
$\Gamma_{htt}^{h}(0,0)$ does not get vector or axial-vector components, so that $C_{3q}$
remains equal to zero\footnote{See Ref.~\cite{Mihaila:2010mp} for details.}
and no mixing between ${\cal O}_{2t}$ and ${\cal O}_{3t}$ occurs at the given order.

 Moreover, we also proved that 
\begin{eqnarray}
\zeta_v &=&\frac{\zeta_{m_t}}{\zeta_{y_t}}\,,
\end{eqnarray}
where $\zeta_{y_t}$ and $\zeta_{m_t}$ were defined in Eqs.~(\ref{eq::letg}) and (\ref{eq::letse}). The
decoupling coefficient for the vacuum expectation value ${\zeta_v}$ can be
derived from the relation $M_W=g_2 v/2$. At ${\cal
  O}(\alpha_t)$ and ${\cal  O}(\alpha_s \alpha_t)$  only 
the transversal part of the $W$-boson propagator contributes, because 
the decoupling coefficient of the gauge coupling $g_2$  receives only corrections
of ${\cal O}(\alpha_2)$ or higher. Therefore, at this level of accuracy holds
\begin{eqnarray}
\zeta_{v} &=&1+\frac{\Pi_W^{T,h}(0)}{2 M_W^2}\,.
\end{eqnarray}

As can be understood from the analytical expression displayed above, the
decoupling coefficients for $\alpha_s, \alpha_t$ and $m_t$ have a remnant
logarithmic dependence on the  new physics 
scale (in our case it is identified with $M_{\rm SUSY}$) and a polynomial
dependence on the mixing parameters even for the case that $M_{\rm SUSY}\gg
M_{\rm EW}$. These corrections we denote generically as non decoupling effects,
in the sense that they do not vanish when the scale of new physics become much
heavier than the electroweak one. Therefore, the energy evolution of the three fundamental
parameters   contain important informations   about the underlying theory,
even in the challenging case of the decoupling scenario.

%\begin{itemize}
%\item 1-loop + ${\cal O} (\epsilon)$ (??) ; complete;   agreement with arxiv:1407.4081
%\item 2-loop results for two limits: i) all susy particles have the same mass
%  $M_{\rm SUSY}$
%  and ii) low $\mu_{\rm SUSY}$, i.e. $\mu_{\rm SUSY} \ll M_{\rm SUSY}$ 
%\end{itemize}

\section{Numerics} 
In this section, we study the phenomenological implications of the two-loop
calculation presented above. For our analysis we consider  the strong
coupling constant, the top-quark mass and the top-Yukawa
coupling. For a detailed study on the bottom sector we refer for example 
 to Refs.~\cite{Noth:2010jy,Mihaila:2010mp}. For the SM input
parameters we employed the following numerical values: $M_W=80.387 \pm 0.016$~GeV, 
$M_t=173.34\pm 0.81$~GeV, $M_h=125.09\pm 0.24$~GeV and
$\alpha_s(M_Z)=0.1181\pm 0.011$~\cite{pdg}. For the SUSY parameters, we focus
on two benchmark scenarios that are still allowed by the direct searches at the LHC.
 The first   scenario is the pMSSM  as defined in
Ref.~\cite{pmssm} and is characterized  by a heavy higgsino 
sector in the TeV range. More precisely, the numerical values of the specific
parameters read $\mu = 2.5$~TeV, $A_t=-4.8$~TeV,
$\tan\beta=10$, $M_A=1.5$~TeV, $M_{\tilde{g}}=1$~TeV, $M_Q\simeq M_U\simeq
M_D\simeq 2.5$~TeV. The second scenario we study is a MSUGRA based scenario
as introduced in Ref.~\cite{msugra}, with heavy SUSY particles up to $6$~TeV
but light electroweakino around $200$~GeV. Explicitly, $m_0=6183$~GeV, $m_{1/2}=470$~GeV, $A_0=-4469$~GeV,
$\tan\beta=10$, $\mu > 0$.

Two-loop decoupling effects are usually combined with three-loop
renormalization group equations (RGEs) in order to properly resum the
logarithms that arise in the calculation. For a detailed discussion of the
running and decoupling procedure within the \msbar{} scheme we refer to
Ref.~\cite{Chetyrkin:2000yt}. The application of the method to the \drbar{}
scheme follows in complete analogy.

 For the current analysis, we
implement the three-loop  RGEs for the gauge and Yukawa couplings from
Refs.~\cite{smbeta} for the SM and from Ref.~\cite{Ferreira:1996ug} for the MSSM. 
 The three-loop anomalous dimension for the top-quark mass is
 not available in the literature neither in the SM nor in the MSSM. However,
 the genuine QCD and SUSY-QCD contributions can be derived from the beta-function of the
 top-Yukawa coupling.   The dominant mixed
 (SUSY)QCD-Yukawa contributions of ${\cal O}(\alpha_s^2\alpha_t,
 \alpha_s\alpha_t^2)$ we have computed explicitly and implemented in our
 numerical analysis. Furthermore, we took into 
 account the electroweak contributions  at two loops in the
 running and at one loop in the decoupling coefficients.  As it is well known,
 when the tadpole contributions are renormalized to zero, there is a gauge
 dependence of the running masses. In our setup, we chose the Feynman-gauge
 for the electroweak sector. Nevertheless, the decoupling coefficients for the
 running masses remain gauge independent. In our numerical analysis, the gauge
 dependence is hidden in the SM value of $m_t(M_t)$ that is our starting value
 for the running analysis. 

\begin{figure}[ht]
  \begin{center}
    \begin{tabular}{cc}
      \hspace*{-1.cm}
      \epsfig{figure=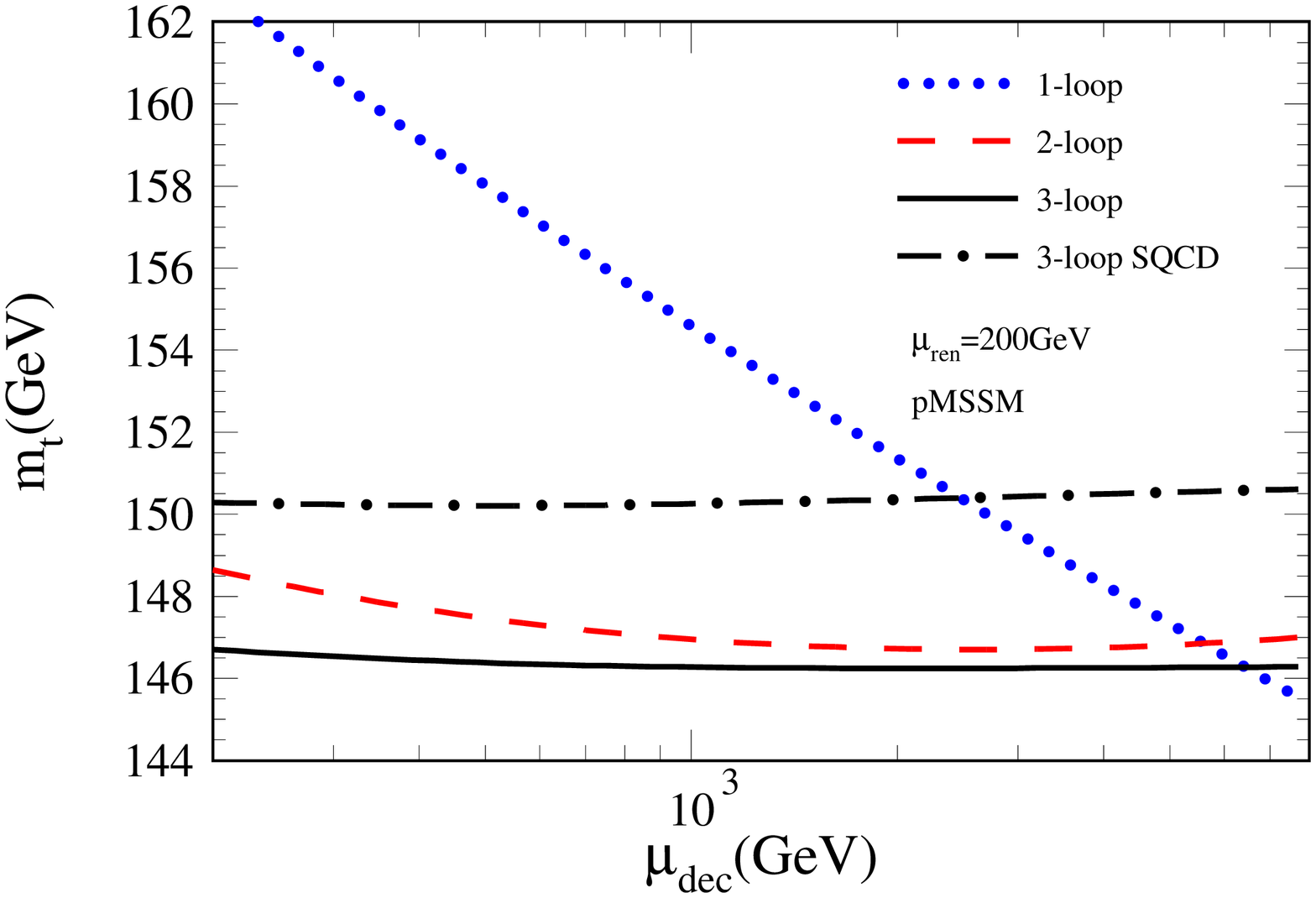,width=22em}
      &
      \hspace*{-1.cm}
      \epsfig{figure=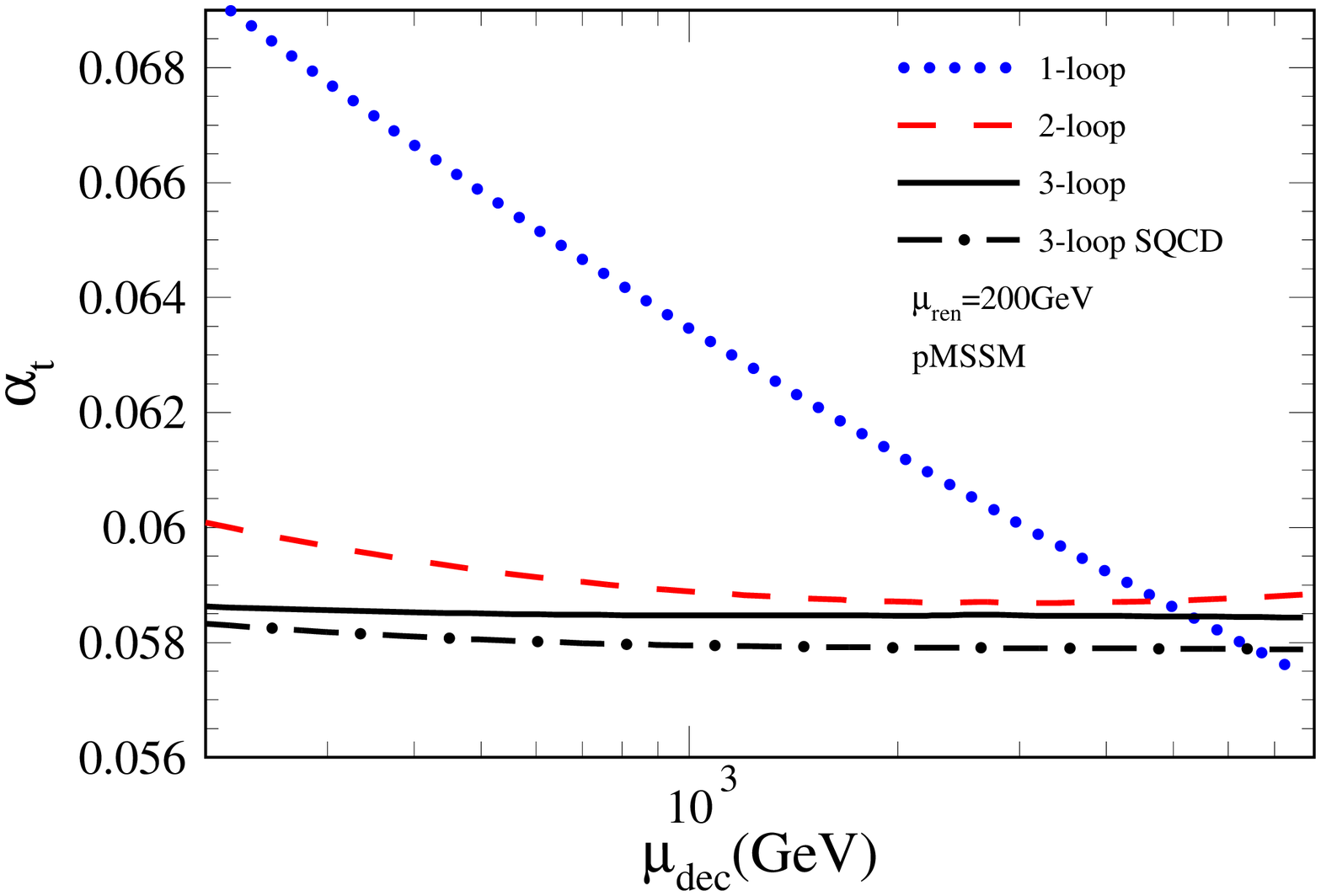,width=22em}
      \\(a) & (b)\\
      \multicolumn{2}{c}{
        \epsfig{figure=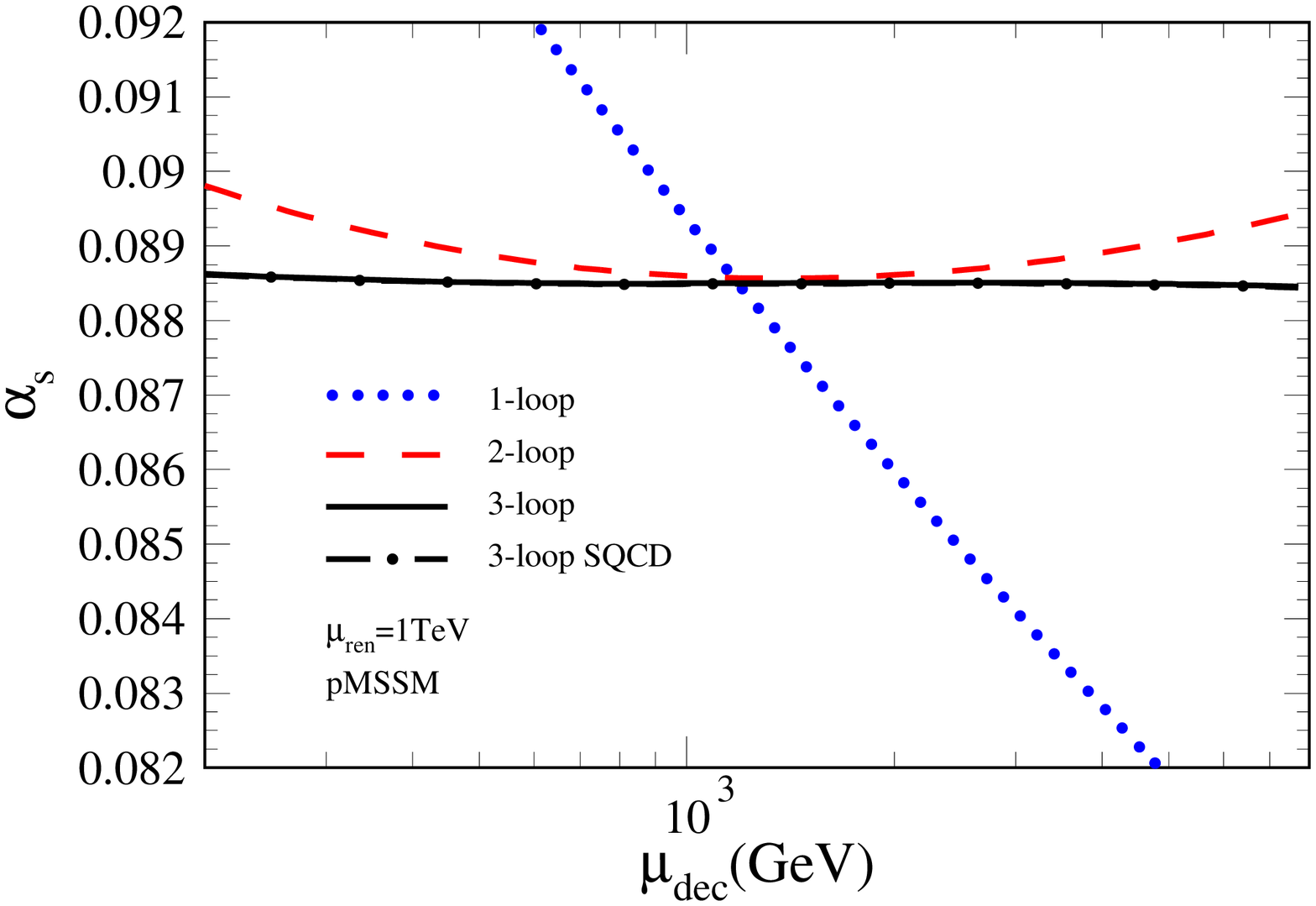,width=22em}
      }
      \\
      \multicolumn{2}{c}{(c)}\\
    \end{tabular}
    \parbox{14.cm}{
      \caption[]{\label{fig::mudec}\sloppy
        $m_t(\mu_{\rm ren}=200\,\mbox{GeV})$, $\alpha_t(\mu_{\rm
          ren}=200\,\mbox{GeV})$, and
        $\alpha_s(\mu_{\rm ren}=1\,\mbox{TeV})$ as  functions 
        of $\mu_{\rm dec}$.
        The dotted, dashed and full lines correspond to
        one-, two- and three-loop running. The dot-dashed lines display the
        three-loop SUSY-QCD contributions.
        }}
  \end{center}
\end{figure}

The choice of the scale $\mu_{\rm dec}$ at which the SM is matched with the MSSM is not fixed
by the theory and any remnant dependence of the physical parameters on it is
a measure of the theoretical uncertainties. In Fig.~\ref{fig::mudec} we show  the 
dependence on this scale for $m_t(\mu_{\rm ren}=200\,\mbox{GeV})$, $\alpha_t(\mu_{\rm
          ren}=200\,\mbox{GeV})$, and $\alpha_s(\mu_{\rm ren}=1\,\mbox{TeV})$
within the pMSSM scenario.   The  dotted (blue), dashed (red)
and full (black) lines in the figure correspond to one-, two- and three-loop
running. The dot-dashed lines stand for the three-loop SUSY-QCD
contributions. Explicitly, to obtain this plot we employed the three-loop RGEs
for QCD and SUSY-QCD and the ${\cal O}(\alpha_s, \alpha_s^2)$ contributions to
the decoupling coefficients. As expected, going from one- to two- and three-loop order 
 the matching scale dependence stabilizes and the three-loop results are
practically independent. It is also interesting to observe that the matching
scale dependence  at two loops exceeds the current experimental uncertainty on
$m_t$ and $\alpha_s$ by about a factor of two.\\
Furthermore, the best choice for
the matching scale (defined as the scale where the higher order radiative
corrections are minimal)
 is different for $m_t$, $\alpha_t$  and $\alpha_s$,
respectively. This feature can be understood from  Eqs.~(\ref{eq:zetas1},\ref{eq:zetat1},\ref{eq:zetam1}) and
(\ref{eq:zetas2},\ref{eq:zetat2},\ref{eq:zetam2}). For
 $\alpha_s$ the best choice of $\mu_{\rm dec}$ is approximately given by the average mass
of the coloured SUSY particles. In the top sector, however, the logarithmic
dependence  on the mass parameters is supplemented by the terms proportional
with $\tilde{X}_t,\tilde{X}_t^2,\tilde{X}_t^3$. For the pMSSM scenario
$\tilde{X}_t\simeq 2$ and the scale at which the radiative 
corrections vanish increases towards   $6$~TeV. For small $X_t$ values, {\it e.g.}
for small trilinear coupling $A_t$ and large values for $\tan \beta$, the
terms proportional with powers of $\tilde{X}_t$ will drop off and
$\zeta_{m_t}$ and $\zeta_{y_t}$ will  have almost a logarithmic
dependence on the masses of the supersymmetric particles.

In principle, one can perform
the decoupling procedure at each mass threshold. This approach guarantees the
absence of potentially large logarithmic and power corrections.
 However, a tower of
intermediate non supersymmetric  effective theories with  a complicated mixing
pattern will arise.  In our framework we avoid these computational
complications by decoupling all heavy particles in one step. Nevertheless,
 in order to reduce the dependence of the theoretical predictions  on the
  matching scale, we have to  take into
 account  higher order radiative corrections. Fig.~\ref{fig::mudec} 
 demonstrates the necessity for the two loop
 corrections to the decoupling coefficients and their phenomenological
 implications.

\begin{figure}[ht]
  \begin{center}
    \begin{tabular}{cc}
      \hspace*{-1cm}
      \epsfig{figure=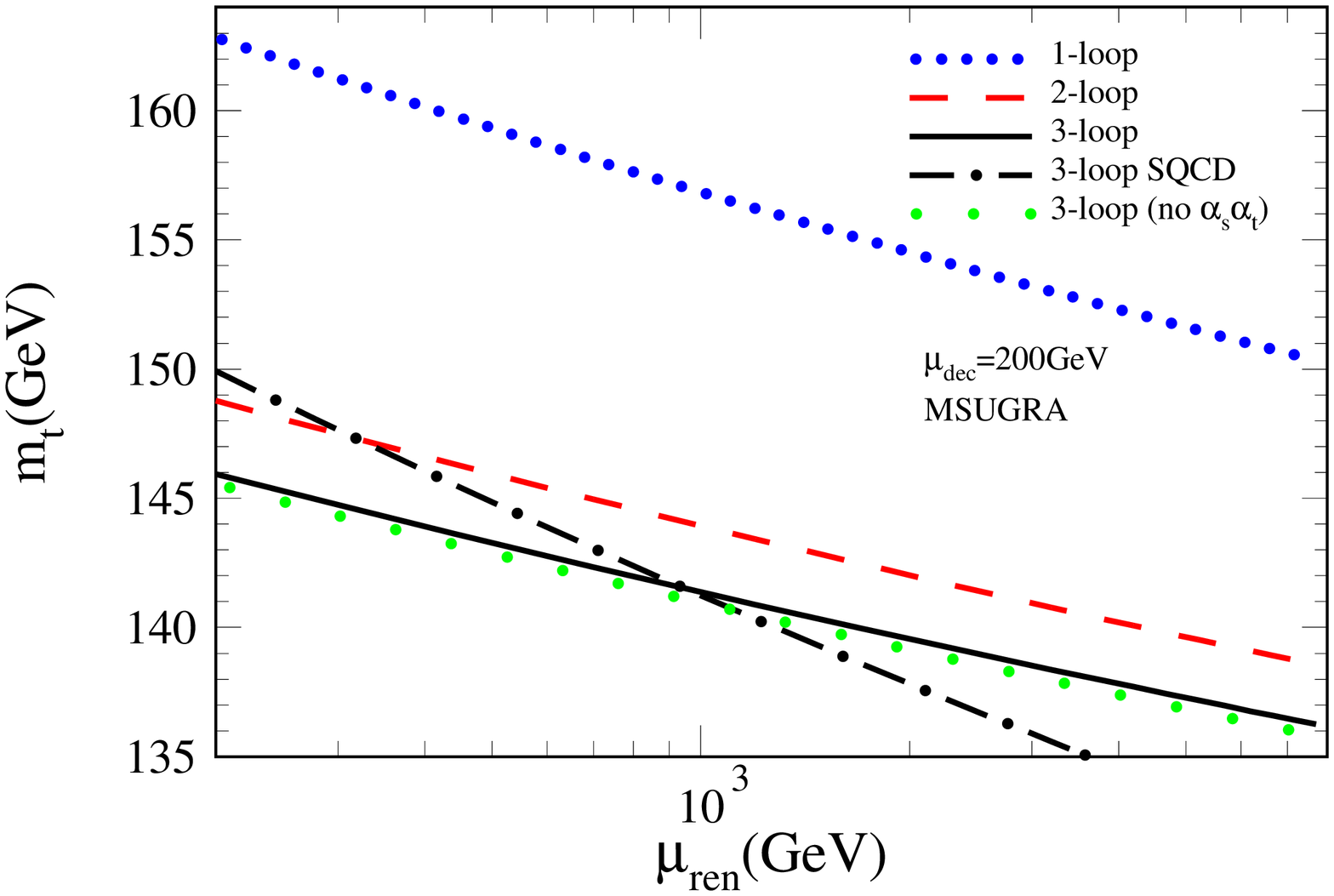 ,width=22em}
      &
      \hspace*{-1cm}
      \epsfig{figure=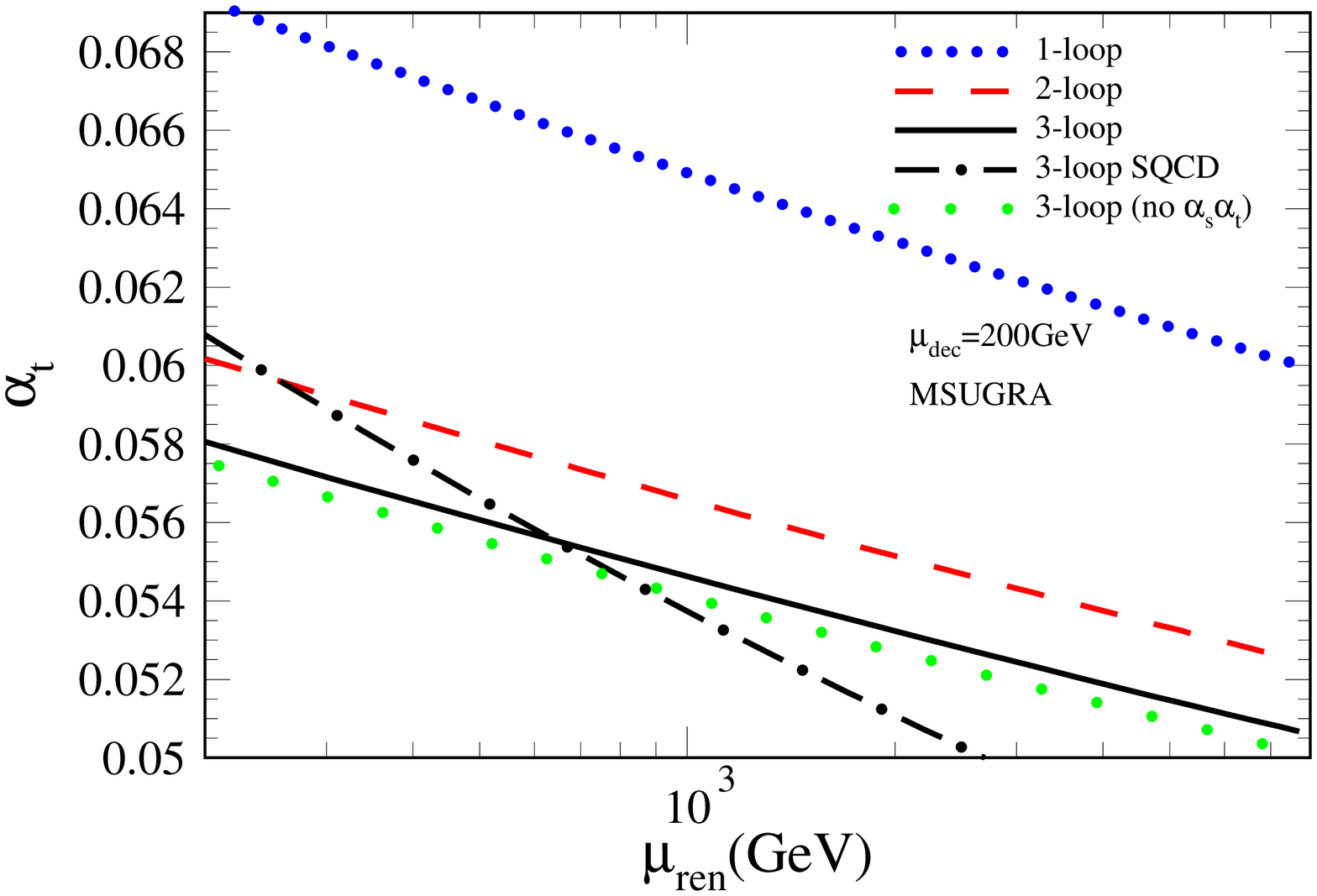,width=22em}
      \\(a) & (b)\\
      \multicolumn{2}{c}{
        \epsfig{figure=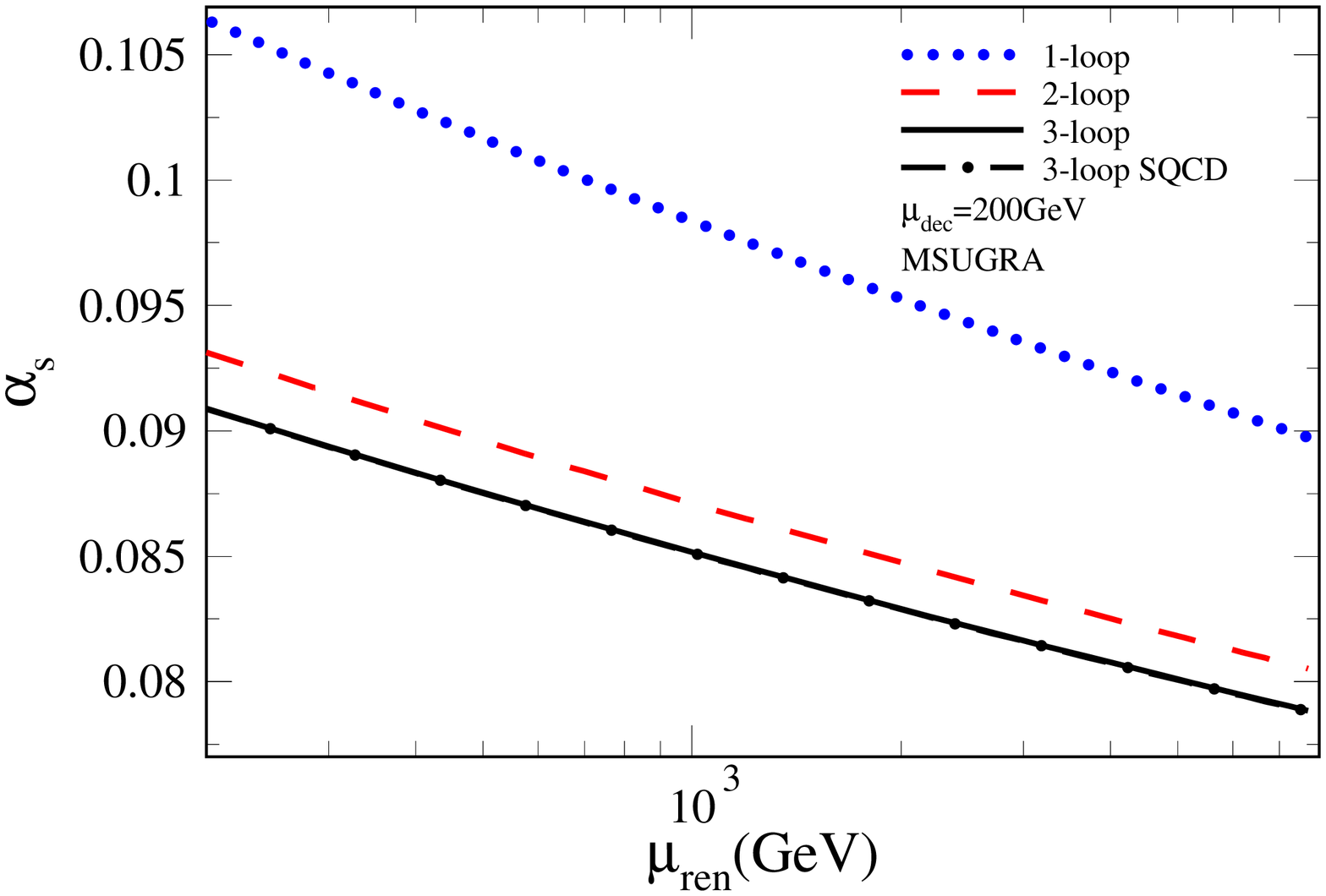,width=22em}
      }
      \\
      \multicolumn{2}{c}{(c)}\\
    \end{tabular}
    \parbox{14.cm}{
      \caption[]{\label{fig::muren1}\sloppy Renormalization scale dependence
        for 
        $m_t$, $\alpha_t$, and $\alpha_s$ for  $\mu_{\rm dec}=200$~GeV.
        The dotted, dashed and full lines correspond to
        one-, two- and three-loop running. The dot-dashed lines display the
        three-loop SUSY-QCD contributions. The large dotted lines marked in
        the legend  with the label (no $\alpha_s\alpha_t$ ) show the
        predictions of the three-loop analysis, where the ${\cal
          O}(\alpha_s\alpha_t)$ contributions to the decoupling coefficients
        for $m_t$ and $\alpha_t$ were excluded. 
        }}
  \end{center}
\end{figure}

\begin{figure}[ht]
  \begin{center}
    \begin{tabular}{cc}
      \hspace*{-1cm}
      \epsfig{figure=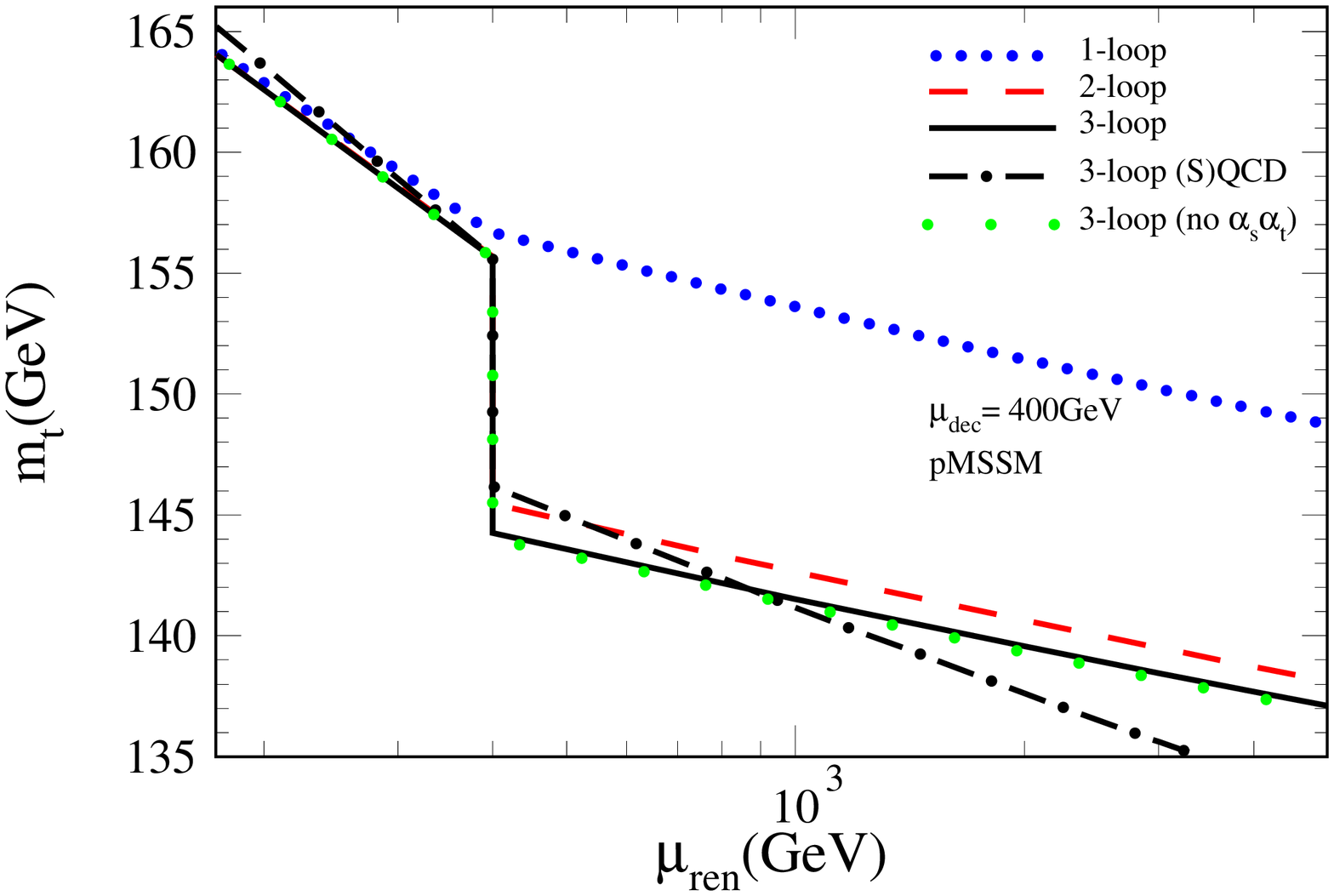 ,width=22em}
      &
      \hspace*{-1cm}
      \epsfig{figure=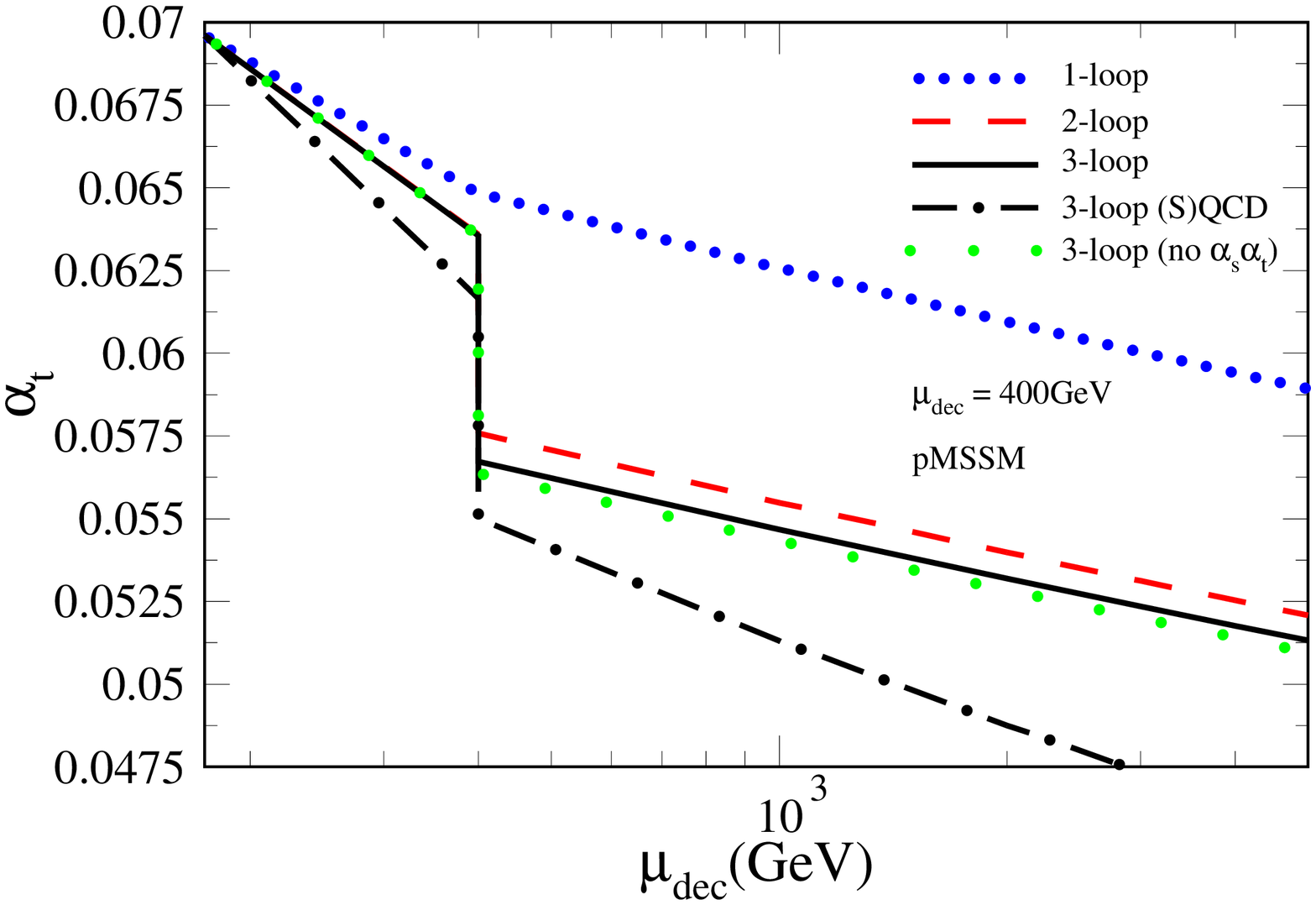,width=22em}
      \\(a) & (b)\\
      \multicolumn{2}{c}{
        \epsfig{figure=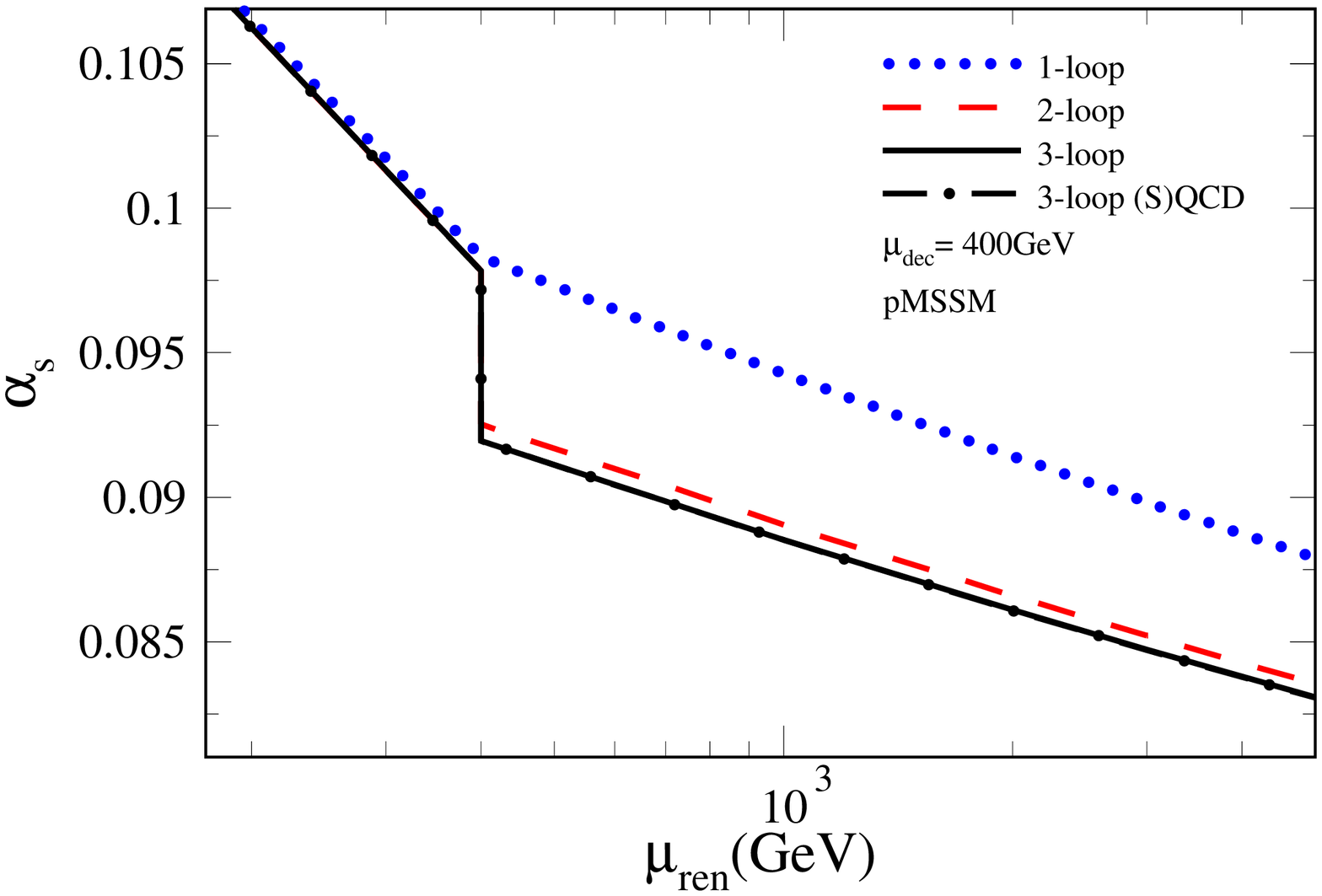,width=22em}
      }
      \\
      \multicolumn{2}{c}{(c)}\\
    \end{tabular}
    \parbox{14.cm}{
      \caption[]{\label{fig::muren2}\sloppy Running of
        $m_t$, $\alpha_t$, and $\alpha_s$.  The matching scale between the SM
        and the MSSM is $\mu_{\rm dec}=400$~GeV. The convention for the lines
        is as in the previous plot. The jumps at $\mu_{\rm dec}=400$ display
        the decoupling effects.
        }}
  \end{center}
\end{figure}

 The top-Yukawa contributions are of phenomenological relevance only
for $m_t$ and $\alpha_t$, as can be read from the difference between the full
and dot-dashed lines in Fig.~\ref{fig::mudec}. For example, the top-Yukawa
contributions to  $m_t(\mu_{\rm ren}=200\,\mbox{GeV})$ 
amounts to about $4$~GeV, a value almost four times larger than the same
quantity within the SM. The $4$~GeV in the running mass in the MSSM can be
explained through the top-Yukawa effects  both on the running and on the
decoupling.  This aspect is also nicely illustrated in
Fig.~\ref{fig::muren2}(a), where for the chosen decoupling scale
of $\mu_{\rm dec}=400$~GeV
 the dominant effects are induced by the modifications in the RGEs  due to
 Yukawa couplings.
It is also interesting to note that the relative size between SUSY-QCD and 
top-Yukawa contributions varies with the choice of the renormalization
scale, as can be understood from  Fig.~\ref{fig::muren1}.
For the $\alpha_s$, however, the  top-Yukawa contributions are not
phenomenologically relevant, as can be understood from the superposition of
the dot-dashed and full lines in the lower plot of Fig.~\ref{fig::mudec}.

In Fig.~\ref{fig::muren1}  the renormalization scale dependence
of $m_t$, $\alpha_t$ and $\alpha_s$ within the MSUGRA scenario is shown, where
the matching scale 
was fixed at $\mu_{\rm dec}=200$~GeV. The convention for the lines is the same
as in the previous figure up to the large dotted lines that display  the
three-loop analysis, for which the ${\cal O}(\alpha_s\alpha_t)$ 
contributions to the decoupling coefficients were not taken into account. 
 Let us mention that the size of the
three-loop contributions  is few times larger than the current experimental
uncertainties for $m_t$ and $\alpha_s$~\cite{pdg}  and comparable with the
expected 
accuracy on $\alpha_t$ at future colliders. As can be understood from
comparing the full and dot-dashed lines in the figure, the  genuine SUSY-QCD
contributions are sufficient for the prediction of the energy evolution of
$\alpha_s$. However, the  running of $m_t$ and $\alpha_t$ receive significant
corrections from the Yukawa and/or mixed QCD-Yukawa sectors, that
can well exceed the current experimental accuracy  at high energy scales.
Let us also mention that the  ${\cal O} (\alpha_t,\alpha_s\alpha_t)$
  corrections 
  to the running and threshold effects within the MSSM 
are few times larger than  in the SM. This behaviour can be
explained by the interplay between the masses and the mixing parameters of the
model. Moreover, as can be understood from the comparison of the solid and
large dotted lines, the contributions of ${\cal O}(\alpha_s\alpha_t)$ to the
decoupling coefficients  have a small numerical effect, well below the
experimental accuracy.

For a better understanding of the phenomenological implications we show in
Fig.~\ref{fig::muren2} the scale evolution of the three parameters both in
the SM and the MSSM  for the pMSSM scenario, where the matching between the
two theories was performed at $\mu_{\rm dec}=400$~GeV. Namely, below $400$~GeV
the SM is considered as the theory describing the physical phenomena. Above
this energy scale, the MSSM is the underlying theory. For the dot-dashed
lines, only the QCD (below $\mu_{\rm dec}=400$~GeV) and SUSY-QCD (above $\mu_{\rm
  dec}=400$~GeV) contributions are taken into account. 
  For the one-loop
running (dotted lines in the plots) only a change of slopes occurs, because no
decoupling is taken into account. At two and three loops, the running is
accompanied by one- and two-loop decoupling. The vertical shifts at
$\mu_{\rm ren}=\mu_{\rm dec}=400$~GeV display directly the effects of the
decoupling procedure. The numerical values of the mixed (SUSY)QCD-Yukawa
contributions at three loops can be read from the difference
between the full and the dot-dashed curves in the figure.  As already pointed
out, their magnitude depends on the scale choice and are
phenomenologically significant for the top sector of the MSSM.   The
magnitude of the ${\cal O}(\alpha_s\alpha_t)$ corrections to the decoupling
coefficients for $m_t$ and $\alpha_t$ can be read from the difference between
the solid and the large dotted lines. For the  MSSM scenario under
consideration, they are much smaller than the expected experimental precision.

\section{Conclusions}
 In this paper we have calculated two loop mixed QCD-Yukawa corrections to the
 decoupling coefficients for the strong coupling, the top-Yukawa coupling and
 the top quark mass. As underlying theory we considered the MSSM and
 decoupled all the supersymmetric particles in one step. The price for this
 simplifying assumption is the necessity for  two-loop corrections to the
 matching coefficients for the fundamental parameters like coupling constants
 and particle masses. The two-loop decoupling coefficients together with the
 three-loop RGEs ensure the independence of the decoupling procedure on the
 scale at which the heavy particles are integrated out, up to higher order
 terms in $\alpha_s$ and $\alpha_t$. 
             
 Although the running masses and couplings are themselves
 not physical observables, they are necessary ingredients for the theoretical
 predictions  of cross sections, branching ratios or physical
 masses. For example, if in a  diagrammatic calculation performed within the
 MSSM at fixed order  or in the  EFT approach the
 running top-quark 
 mass and/or running  top-Yukawa and strong couplings are used,
 then the non-decoupling effects  we discussed here are implicitly contained
 in the numerical values of the running parameters. Depending on the order in
 perturbation theory at which the calculation is performed, the ${\cal
   O}(\alpha_s^2,\alpha_s\alpha_t)$ might not be required. In this case the
 decoupling scale have to be carefully chosen in order to avoid missing large
 higher order radiative corrections to the running parameters as we have shown
 in Fig.~\ref{fig::mudec}.\\ 
There is also another method to  determine the numerical
 values of the running top-quark mass and top-Yukawa coupling. Namely, they
 can  be  derived  from  the measured top, $W$ and $Z$ boson, and Higgs pole
 masses. However, such  a determination will 
 get very large radiative corrections for heavy supersymmetric
 particle~\cite{Kunz:2014gya}. 
 Fig.~3 and Fig.~5 of Ref.~\cite{Kunz:2014gya}  display  the explicit
 comparison of the method based on running and
 decoupling procedure (similar with  that presented  in the current paper) and
 the direct calculation starting from the relation
 between the pole and the running top quark mass  within the SUSY-QCD,
 using the code
 {\tt TSIL}~\cite{Martin:2005qm}. It turned out that, in order to
bring the theoretical uncertainties on the running top quark mass at least  in
the range of the experimental 
precision, one needs  even the three-loop corrections. Such type of
calculations are   computationally very involved
and not yet available in the literature. For
the mixed (SUSY)QCD-Yukawa corrections to these relations,
  we expect a similar behaviour in the perturbative
 expansion but being technically even more involved. 
However, from the numerical analysis we presented in the previous section, one
can see that the difference  between   the two- and the three-loop corrections
derived within the running and decoupling approach are significantly
smaller. For example the genuine three-loop  contributions  to the running
top-quark mass are of about                
  $1$~GeV for appropriately chosen decoupling scales, as can be read from
Fig.~\ref{fig::muren2}(a). 
 This behaviour can be explained by the fact that the large logarithms of the
 form 
 $\ln(M_{\rm top}^2/M_{\rm SUSY}^2)$ are resummed through the use of the RGEs
 in the SM and the MSSM. In our setup, we use the relation between the pole
 and the running masses and couplings only within the SM, so that the occurring
   logarithms  of the  form $\ln(M_{\rm top}^2/M_{\rm EW}^2)$, where
   $M_{\rm   EW}=M_{\rm Z},M_{\rm W}$ and $M_{\rm h}$, are numerically small.

  From our numerical analysis it turned
 out that the genuine ${\cal O}(\alpha_s\alpha_t)$ contributions to the
 matching  coefficients are well below the experimental precision. However,
 the complete  ${\cal O}(\alpha_t,\alpha_s\alpha_t)$ corrections  (including
 the RGE  and the threshold effects) to the running
 top quark mass amount to about few GeV and for the top-Yukawa coupling reach
 the percent range. The ${\cal O}(\alpha_s\alpha_t)$   corrections to the
 decoupling coefficients for $m_t$, $\alpha_s$ and $\alpha_t$   are
 necessary, for example, for the   prediction of the lightest Higgs boson
 mass in   supersymmetric theories and for the vacuum stability studies in such
  theories, when going beyond two-loop accuracy. 
 %For example, a reduction of the running top quark mass of about
 % $3$~GeV as shown in Fig.~\ref{fig::mudec} will imply a decrease of the
 % predicted light Higgs boson mass by the 
 % same amount. 
  The latter are just two examples of current analyses that play key roles in
  constraining and/or 
  unrevealing the type and scale of new physics.  
% However, we leave the  phenomenological study for a future dedicated paper.

Furthermore, we provide compact analytical formulae for the decoupling
coefficients for two mass hierarchies: i) a completely degenerate
supersymmetric mass spectrum
and  ii) a quasi degenerate supersymmetric mass spectrum with higgsinos much
lighter than the rest of the superpartners. Along with this paper we provide 
the results in electronic form, that should be useful for other calculations.

\vspace*{1em}

\noindent
{\bf Acknowledgments}\\
This work was supported by  {\it Deutsche Forschungsgemeinschaft} (contract
MI~1358,  Heisenberg program). The authors are grateful to M. Steinhauser for
collaboration at an early stage of the project, for numerous enlightening
discussions and for reading the manuscript.

\appendix
\section{Appendix A} 
\label{App:AppendixA}

In this appendix we provide the complete one-loop results for the decoupling
coefficient of top-quark mass. It reads

\begin{align}
\zeta_{m_t}=&1 
 + a_s \Bigg\{
C_R x_{\rm DRED} 
 + C_R \Bigg[
\frac{1-2  x_{\tilde{g} Q_3}^2}{2  x_{\tilde{g} Q_3}^2} \mathcal{P}_{\tilde{g}Q_3} 
  -\frac{1}{2} \mathcal{P}_{\tilde{g}U_3} 
 + \Big(
-\frac{1}{2} \mathcal{P}_{\tilde{g}Q_3}^2 
  -\frac{1}{2} \mathcal{P}_{\tilde{g}U_3}^2\Big) L_{\tilde{g}}
\nonumber
\\& \displaybreak[1]
 + \Big(
-\frac{1}{2} 
 + \frac{1}{2} \mathcal{P}_{\tilde{g}Q_3}^2\Big) L_{Q_3}
 + \Big(
-\frac{1}{2} 
 + \frac{1}{2} \mathcal{P}_{\tilde{g}U_3}^2\Big) L_{U_3}
 + \frac{X_t}{m_{Q_3}} \Big[
\frac{2}{ x_{\tilde{g} Q_3}} \mathcal{P}_{\tilde{g}Q_3} \mathcal{P}_{\tilde{g}U_3} L_{\tilde{g}}
\nonumber
\\&  
 -2 x_{\tilde{g} Q_3}\mathcal{P}_{\tilde{g}Q_3} \mathcal{P}_{Q_3U_3} L_{Q_3}
 + \mathcal{P}_{\tilde{g}Q_3} \Big(
-\frac{2}{ x_{\tilde{g} Q_3}} \mathcal{P}_{\tilde{g}U_3} 
 + 2 x_{\tilde{g} Q_3} \mathcal{P}_{Q_3U_3}\Big) L_{U_3}\Big]\Bigg]\Bigg\}\nonumber\displaybreak[1]
\\&  \displaybreak[1]
 + a_2 \Bigg[
x_{\rm DRED} \Big(
-\frac{3}{8 }
  \Big)  
-\frac{3}{16} 
  -\frac{3}{8} \mathcal{P}_{2Q_3} 
 + L_{2} -\frac{3}{8} \mathcal{P}_{2Q_3}^2 
 + \Big(
-\frac{3}{8} 
 + \frac{3}{8} \mathcal{P}_{2Q_3}^2\Big) L_{Q_3}\Bigg] \nonumber
\\&
 +  a_1 \Bigg\{x_{\rm DRED} \Big(
  -\frac{1}{72}\Big)+
\frac{17-9  x_{1 Q_3}^2}{144 x_{1 Q_3}^2} \mathcal{P}_{1Q_3} 
  -\frac{2}{9} \mathcal{P}_{1U_3} 
 + \Big(
-\frac{1}{72} \mathcal{P}_{1Q_3}^2 
  -\frac{2}{9} \mathcal{P}_{1U_3}^2\Big) L_{1}
\nonumber
\\&\displaybreak[1]
 + \Big(
-\frac{1}{72} 
 + \frac{1}{72} \mathcal{P}_{1Q_3}^2\Big) L_{Q_3}
 + \Big(
-\frac{2}{9} 
 + \frac{2}{9} \mathcal{P}_{1U_3}^2\Big) L_{U_3}
 + \frac{X_t}{m_{Q_3}} \Big[
\frac{2}{9  x_{1 Q_3}} \mathcal{P}_{1Q_3} \mathcal{P}_{1U_3} L_{1}
\nonumber
\\& 
  -\frac{2 x_{1 Q_3}}{9} \mathcal{P}_{1Q_3} \mathcal{P}_{Q_3U_3} L_{Q_3} 
 + \mathcal{P}_{1Q_3} \Big(
-\frac{2}{9 x_{1 Q_3}} \mathcal{P}_{1U_3} 
 + \frac{2 x_{1 Q_3}}{9 } \mathcal{P}_{Q_3U_3}\Big) L_{U_3}\Big]\Bigg\} \nonumber
\\& 
 + a_t \Bigg\{
-\frac{9}{8} 
 + \frac{1}{2} \mathcal{P}_{Q_3\mu} 
 + \frac{1}{4} \mathcal{P}_{U_3\mu} 
 + c_{\beta}^2 \Big(
-\frac{3}{8} 
  -\frac{3}{4} L_{A}\Big) 
 + \Big(
- \mathcal{P}_{Q_3\mu} 
 + \frac{1}{2} \mathcal{P}_{Q_3\mu}^2\Big) L_{Q_3}\nonumber
\\& 
 + \Big(
-\frac{1}{2} \mathcal{P}_{U_3\mu} 
 + \frac{1}{4} \mathcal{P}_{U_3\mu}^2\Big) L_{U_3}
 + \Big(
-\frac{3}{4} 
 + \mathcal{P}_{Q_3\mu} 
  -\frac{1}{2} \mathcal{P}_{Q_3\mu}^2 
 + \frac{1}{2} \mathcal{P}_{U_3\mu} 
  -\frac{1}{4} \mathcal{P}_{U_3\mu}^2\Big) L_{\mu}\Bigg\}\nonumber
\\\displaybreak[1]& 
 + a_b \Bigg\{
-\frac{1}{2} 
 + \frac{1}{4} \mathcal{P}_{D_3\mu} 
  -\frac{1}{4} L_{A}
 + c_{\beta}^2 \Big(
-\frac{7}{8} 
  -\frac{3}{4} L_{A}\Big) 
 + \Big(
-\frac{1}{2} \mathcal{P}_{D_3\mu} 
 + \frac{1}{4} \mathcal{P}_{D_3\mu}^2\Big) L_{D_3}\nonumber
\\& 
 + \Big(
-\frac{1}{4} 
 + \frac{1}{2} \mathcal{P}_{D_3\mu} 
  -\frac{1}{4} \mathcal{P}_{D_3\mu}^2\Big) L_{\mu}
 + 
\frac{X_b}{\mu\, t_{\beta}} \Bigg[
- \mathcal{P}_{D_3\mu} \mathcal{P}_{D_3 Q_3} L_{D_3}
 + \mathcal{P}_{D_3 Q_3} \mathcal{P}_{Q_3\mu} L_{Q_3}\nonumber
\\& 
 + \Big(
 \mathcal{P}_{D_3\mu} \mathcal{P}_{D_3 Q_3} 
  - \mathcal{P}_{D_3 Q_3} \mathcal{P}_{Q_3\mu}\Big) L_{\mu}\Bigg]\Bigg\}\displaybreak[1]\,,\end{align}\\

where we have used the following notations
\begin{eqnarray*}
L_i&=&\ln(\mu_{\rm dec}^2/m^2_i)\,\quad x_{ij}=\frac{m_i}{m_j}\,,\\
\mathcal{P}_{ij}&= &\frac{m_i^2}{m_i^2-m_j^2}=\frac{1}{1-\frac{m_j^2}{m_i^2}}=\frac{1}{1-x_{ji}^2}\,.\\
\end{eqnarray*}
Here $M_{Q_3}$,$M_{U_3}$,$M_{D_3}$  are the soft
SUSY-breaking parameters of the stop- and sbottom-sector,  $M_1\,,M_2$ and $M_{\tilde{
  g}}$ denotes  the gaugino masses.  Here we adopted the SU(5) normalization for the gauge coupling
$\alpha_1$. The label $x_{\rm DRED}$ marks the contributions induced by the
change from DREG to DRED. The results for $\zeta_{m_b}$ and $\zeta_{y_b}$ can be derived via
the following replacements $t\leftrightarrow b$ and $U_3 \leftrightarrow D_3$.

%- }}}

\end{document}